\documentclass{article}

\usepackage{PRIMEarxiv}

\usepackage[utf8]{inputenc} 
\usepackage[T1]{fontenc}    
\usepackage{hyperref}       
\usepackage{url}            
\usepackage{booktabs}       
\usepackage{amsfonts}       
\usepackage{nicefrac}       
\usepackage{microtype}      
\usepackage{lipsum}
\usepackage{fancyhdr}       
\usepackage{graphicx}       
\graphicspath{{media/}}     
\usepackage{amsmath}
\usepackage{numprint} 
\usepackage{natbib}
\usepackage{array}
\usepackage{caption}
\usepackage{subcaption}
\usepackage{multirow}

\newcommand{\latwavr}[1][]{%
    \frac{1}{|\mathbb{G}|}\sum\limits_{v_i\in\mathbb{G}}w_{\phi} \left( #1 \right)
}

\title{Deep Learning Weather Models for Subregional Ocean Forecasting: A Case Study on the Canary Current Upwelling System}

\author{
  \begin{minipage}[t]{0.32\textwidth}
    \centering
    Giovanny A. Cuervo-Londoño\\
    {\normalfont Oceanografía Física y Geofísica Aplicada (OFYGA), ECOAQUA\\
    Universidad de Las Palmas de Gran Canaria\\
    Las Palmas, Spain\\
    \texttt{giovanny.cuervo101@alu.ulpgc.es}}
  \end{minipage}
  \hfill
  \begin{minipage}[t]{0.32\textwidth}
    \centering
    Javier Sánchez\\
    {\normalfont Centro de Tecnologías de la Imagen (CTIM), IUCES\\
    Universidad de Las Palmas de Gran Canaria\\
    Las Palmas, Spain\\
    \texttt{jsanchez@ulpgc.es}}
  \end{minipage}
  \hfill
  \begin{minipage}[t]{0.32\textwidth}
    \centering
    Ángel Rodríguez-Santana\\
    {\normalfont Oceanografía Física y Geofísica Aplicada (OFYGA), ECOAQUA\\
    Universidad de Las Palmas de Gran Canaria\\
    Las Palmas, Spain\\
    \texttt{angel.santana@ulpgc.es}}
  \end{minipage}
}

\begin{document}
\maketitle

\begin{abstract}
Oceanographic forecasting impacts various sectors of society by supporting environmental conservation and economic activities. Based on global circulation models, traditional forecasting methods are computationally expensive and slow, limiting their ability to provide rapid forecasts. Recent advances in deep learning offer faster and more accurate predictions, although these data-driven models are often trained with global data from numerical simulations, which may not reflect reality. The emergence of such models presents great potential for improving ocean prediction at a subregional domain. However, their ability to predict fine-scale ocean processes, like mesoscale structures, remains largely unknown. This work aims to adapt a graph neural network initially developed for global weather forecasting to improve subregional ocean prediction, specifically focusing on the Canary Current upwelling system. The model is trained with satellite data and compared to state-of-the-art physical ocean models to assess its performance in capturing ocean dynamics. Our results show that the deep learning model surpasses traditional methods in precision despite some challenges in upwelling areas. It demonstrated superior performance in reducing RMSE errors compared to ConvLSTM and the GLORYS reanalysis, particularly in regions with complex oceanic dynamics such as Cape Ghir, Cape Bojador, and Cape Blanc. The model achieved improvements of up to 26.5\% relative to ConvLSTM and error reductions of up to 76\% in 5-day forecasts compared to the GLORYS reanalysis at these critical locations, highlighting its enhanced capability to capture spatial variability and improve predictive accuracy in complex areas. These findings suggest the viability of adapting meteorological data-driven models for improving subregional medium-term ocean forecasting. This also demonstrates the superior flexibility of graph neural networks compared to traditional models, as they can be adapted to new prediction tasks even when originally developed for different purposes.
\end{abstract}


\section{Introduction}
Oceanographic prediction is crucial for understanding climate change and supporting sectors like maritime transport, fisheries, and natural disaster management~\citep{GODAE}. It relies heavily on accurately forecasting mesoscale processes due to their environmental and economic impacts. These processes give rise to distinct structures, influencing mean currents and transporting key ocean properties~\citep{falkowski1991role}. Despite its importance, predicting mesoscale processes remains challenging for operational forecasts ~\citep{Treguier2017ModelingAF, Mourre2018AssessmentOH}, as evidenced by the difficulty in forecasting the Gulf of Mexico Loop Current eddy during the 2010 Deepwater Horizon oil spill~\citep{GulfMexico, liu2013monitoring}.

Traditional oceanographic prediction techniques rely on numerical models that solve physics-based equations. While these models have significantly advanced ocean forecasting, the theoretical foundation—quasi-geostrophic (QG) theory—and the numerical models have inherent limitations that hinder their accuracy in certain contexts. 

Although the QG theory initially advanced our understanding of mesoscale dynamics and ocean prediction, it struggles to address strong currents, cannot account for bathymetric features, fails to incorporate surface density gradients, and does not model frontal dynamics~\citep{WestwardMotionofMesoscaleEddies}. Despite these theoretical limitations, global ocean circulation models remain the primary tools for oceanographic forecasting, although they contribute to the persistent challenges of operational mesoscale prediction.

On the other hand, numerical ocean prediction (NOP) models—such as NEMO~\citep{NEMO}, which forms the core of reanalysis systems like GLORYS~\citep{GLORYS12} and operational forecast systems like PSY4V3R1~\citep{os-14-1093-2018}—remain the current standard for short-term deterministic forecasting. Nevertheless, they cannot accurately represent reality due to diverse constraints such as incomplete understanding of subgrid-scale parameterizations, poorly known forcing fields, insufficient knowledge of interactions with other Earth system components, and restricted computational resources~\citep{Sommer2018OceanCM}. 

Additionally, incomplete observations with spatiotemporal gaps and the limitations of data assimilation schemes—still in continuous improvement—prevent these models from fully capturing the ocean's state. Furthermore, these models do not fully utilize extensive historical data and lack optimization for modern hardware, such as GPUs, which further reduces their efficiency. These limitations highlight the need for novel approaches to improve ocean prediction capabilities.

In recent years, short-term machine learning weather prediction (MLWP) models have emerged in global atmospheric forecasting, surpassing the efficiency and accuracy of traditional numerical systems~\citep{RiseofDataDriven}. Models such as Pangu Weather~\citep{Pangu}, GraphCast~\citep{lam2023graphcast}, Aurora~\citep{Aurora}, NeuralGCM~\citep{NeuralGCM}, Gencast~\citep{Gencast}, or AIFS~\citep{2024arXiv240601465L} have demonstrated the power of this approach,  significantly reducing inference times and computational costs. By leveraging historical data and focusing on spatiotemporal patterns, MLWP models bypass the constraints of incomplete physical understanding and adapt easily to new prediction tasks without architectural modifications~\citep{gmd-11-3999-2018, Scher2018, Pangu}.

Nevertheless, MLWP models heavily depend on data quality and availability, facing challenges with heterogeneous, sparse, spatially discontinuous, or noisy datasets, which can limit performance in certain contexts. Additionally, the physical consistency of these models deteriorates over long timescales due to spectral bias, which can result in numerical instabilities or unrealistic hallucinations~\citep{2023arXiv230407029C}.

While MLWP models perform well in atmospheric systems, their application to oceanography presents distinct challenges. Unlike atmospheric data, which is abundant, spatially continuous, and less influenced by external factors, oceanographic data is constrained by atmospheric interference, which affects observational accuracy and consistency, spatial discontinuities, predominantly induced by the presence of continental landmasses, which significantly disrupt data continuity and turn the training of these models complex.

Recent advancements in machine learning ocean prediction (MLOP) have led to the development of innovative models at global and regional scales. Xihe~\citep{wangXiHeDataDrivenModel2024}, for instance, has been introduced for global ocean forecasting, while SeaCast~\citep{Seacast} and OceanNet~\citep{chattopadhyay_oceannet_2024} focus on regional applications. Similarly, progress in mesoscale ocean forecasting, including studies on eddy shedding predictability in the Gulf of Mexico~\citep{PredictabilityEddyShedding} and data-driven turbulence forecasting using autoregressive techniques~\citep{DataDrivenForecastTurbulence}, has shown promising results.

The Canary current upwelling system (CCUS), the only eastern boundary upwelling system with islands, presents a unique prediction challenge. In this region, intense mesoscale activity results from the interplay between oceanic features, coastal regimes, and bathymetry, driving strong mesoscale stirring and creating a highly dynamic environment~\citep{SANGRA20092100, ARISTEGUI19941509, BARTON1998455}. These dynamics create a complex environment for sub-regional ocean prediction. State-of-the-art MLOP models have not been applied to medium-term forecasts in this region, highlighting the need for novel approaches.

This work adapts the GraphCast model~\citep{lam2023graphcast} for sub-regional ocean forecasting to evaluate its performance under ocean-specific conditions. Using sea surface temperature (SST) as a case study, we introduce modifications to handle challenges like spatial discontinuities and satellite-based data, including spatially masked loss functions. We assess whether the model can capture mesoscale features such as upwelling, potentially addressing the limitations of traditional ocean models in resolving frontal dynamics and subgrid-scale processes.

In this context, SST is useful for studying the dynamics of mesoscale structures~\citep{HAUSMANN201260}, as eddies, fronts, and filaments create distinct thermal signatures, i.e., warm/cold rotating cores. This study uses the Copernicus Marine Service L4 SST reprocessed product (1982–2020), spanning 39 years, for a subdomain within the IBI region (Iberia, Biscay, Ireland) to train and validate ocean models. The dataset provides daily, gap-free sea surface temperature fields, derived from multiple intercalibrated satellite sources.

In the experimental results, we compare the ocean-adapted GraphCast-based graph neural network (GNN) and ConvLSTM-based model~\citep{2015arXiv150604214S} against NEMO-based reanalysis and forecast products~\citep{GLORYS12, os-14-1093-2018}—specifically GLORYS and PSY4V3R1—in terms of predictive skill, computational efficiency, and suitability for short- to medium-term SST forecasting. We focus on the complex CCUS region, which features strong mesoscale turbulence, persistent upwelling fronts, and energetic eddies—challenges for traditional quasi-geostrophic models. We evaluate how well ML-based models capture complex dynamics over different forecast times and their sensitivity to mesoscale features in challenging areas like oceanic islands and coastal capes.

Additionally, the study assesses the ability of these models to capture seasonal and interannual variability in coastal upwelling systems, where mesoscale processes often dominate. Model sensitivity to observational errors and initial conditions is also analyzed, highlighting their influence on forecast degradation. Finally, the work explores operational implications and proposes strategies to bridge the gap between data-driven approaches and traditional numerical paradigms, emphasizing integration of physical constraints and improving resolution in coastal environments.

Our model's superior performance in SST forecasting is evidenced by $>$76\% and 48\% reductions in RMSE over GLORYS at 5- and 10-day lead times, respectively, and by a computational speed nearly $\sim$100× faster (20-day forecasts in 2.3 minutes vs. 7-day forecasts in 4 hours). However, its sensitivity to initial condition errors—reflected in an 8.3\% increase in days exceeding instrumental error thresholds (2017–2020)—and the presence of triangular artifacts that raise RMSE variability by 15–20\% at larger scales point to key architectural limitations. These findings underscore the need for hybrid designs that combine GraphCast’s spatial precision with ConvLSTM-like stabilization to achieve a better trade-off between accuracy and operational robustness.

Section \ref{se:material_methods} presents the study area of the CCUS and describes the L4 SST dataset used in our experiments. Section~\ref{se:forecast_methods} introduces the predictive approaches we evaluate in this work, including the GLORYS12V1 reanalysis, ConvLSTM architecture, and our proposed graph neural network. 
The experimental results, presented in Section~\ref{se:results}, assess the performance of the models, with particular emphasis on regions exhibiting intense mesoscale activity and the principal capes. Section~\ref{se:discussion} analyzes the strengths and limitations of the methods, addresses systematic errors in coastal versus open-ocean zones, and evaluates model robustness under dynamic oceanic conditions. Finally, Section~\ref{se:conclusion} synthesizes the key contributions of this work, outlines practical implications for operational oceanography, and proposes future research directions.

\section{Area of study and dataset}
\label{se:material_methods}
\subsection{The Canary current upwelling region}
\begin{figure}[!ht]
\begin{center}
\includegraphics[width=\linewidth]{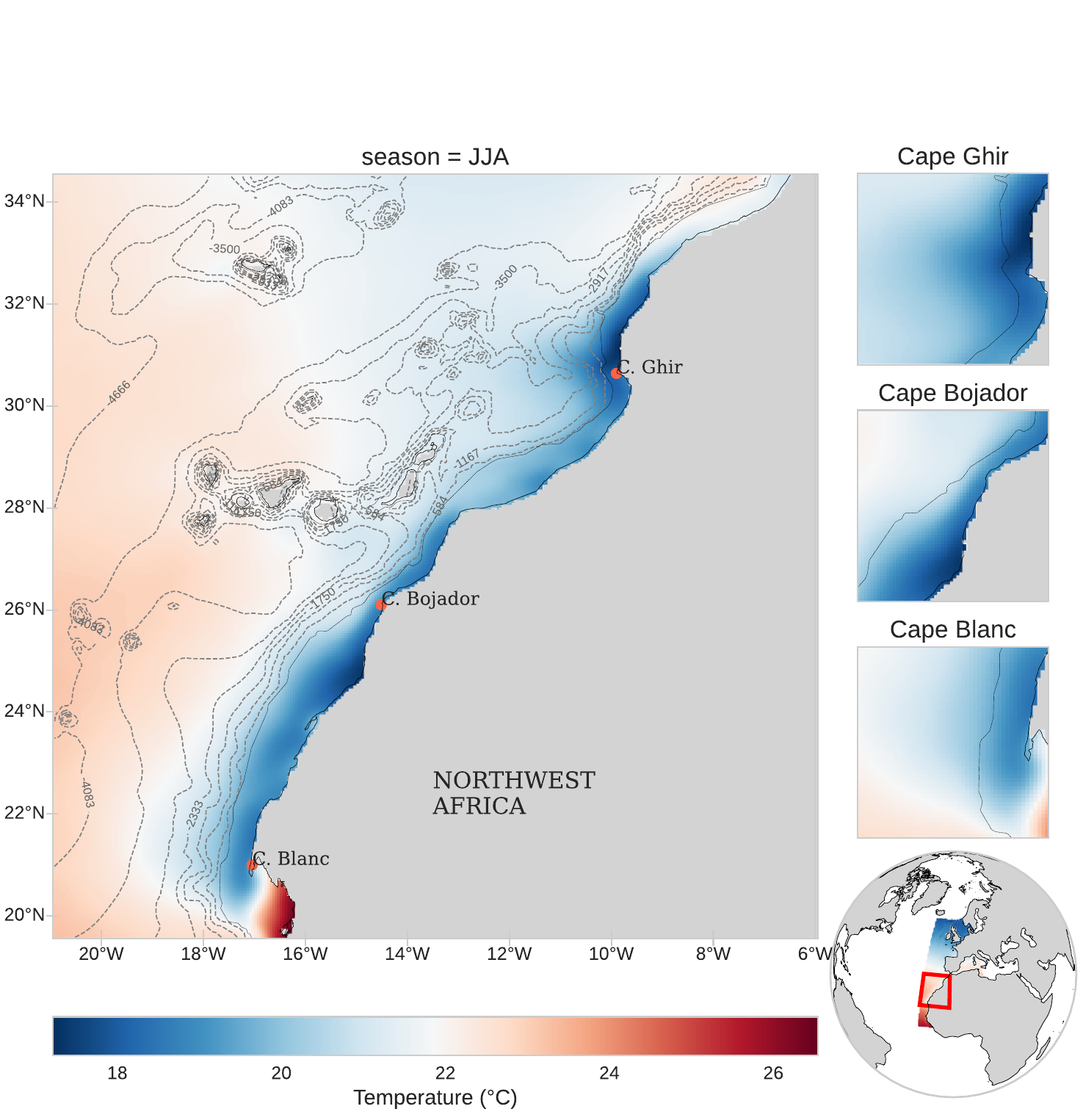}
\end{center}
\caption{Summer (JJA) climatology of sea surface temperature (°C) in the Northwest Africa region, highlighting the coastal upwelling system. The main panel displays the spatial distribution of temperature along the coast, with three prominent upwelling centers: Cape Ghir ($\sim$30°N), Cape Bojador ($\sim$26°N), and Cape Blanc ($\sim$21°N). Dashed gray contours represent isobaths, indicating bathymetric features relevant to upwelling dynamics. Insets on the right provide zoomed views of each cape to better illustrate localized thermal gradients. The bottom-right globe indicates the IBI (Iberian-Biscay-Ireland) domain of the Copernicus dataset, with a red bounding box marking the subdomain corresponding to the study area shown in the main panel.}
\label{fig:bathy_map}
\end{figure}

The study focuses on a subdomain within the Canary Current Large Marine Ecosystem, specifically the Moroccan subregion~\citep{ARISTEGUI200933}, which extends from 21$^\circ$S to 33$^\circ$N between Cape Sim and Cape Blanc; see~\autoref{fig:bathy_map}. This region includes two distinct meridional upwelling zones, as described by~\cite{CROPPER201494}: i) the 21–26$^\circ$N zone, characterized by strong, permanent upwelling throughout the year; and ii) the 26–35$^\circ$N zone, where upwelling remains permanent but is weaker, intensifying during summer due to the seasonal migration of trade winds. The contrast between these two upwelling zones, both in intensity and seasonal variability, highlights the complexity of the Moroccan subregion and underscores its importance for accurately predicting sea surface temperature dynamics.

The region experiences upwelling-favorable winds year-round, with peak intensity during summer due to the northward migration of the Azores High~\citep{wooster1976seasonal}. Additionally, the region exhibits pronounced mesoscale oceanographic variability driven by geographic heterogeneity, including variations in continental shelf width, prominent capes, and perturbations induced by the Canary Islands, which generate filaments and eddies.

The interplay of interconnected physical processes governs the coastal upwelling dynamics off northwest Africa, particularly around Cape Ghir. Prominent capes, such as Ghir, Sim, and Cantin, serve as critical control points where topography and bathymetry modulate atmospheric and oceanic flows. Wind forcing, the primary driver of upwelling, intensifies Ekman transport between Cape Ghir and Cape Sim, injecting positive relative vorticity~\citep{TROUPIN20121} and enhancing the upwelling of cold, nutrient-rich subsurface waters  through shear-driven turbulent mixing~\citep{10.3389/fmars.2023.1113879}. The Cape Ghir Plateau, a submarine projection of the High Atlas orogeny on land, deflects the coastal jet offshore, inducing a potential vorticity imbalance that drives the formation of the characteristic filament~\citep{hagen1996near}.

This filament displays a dual structure: a cold, surface-intensified core with temperature minima and chlorophyll maxima, surrounded by a broader domain of less dense water influenced by anticyclonic interthermocline eddies (ITEs)~\citep{cape_ghir_sangra}. These recurrent ITEs, located north of the filament, strengthen offshore transport through interactions with the upwelling front. Additionally, the irregular bathymetry causes bifurcations in the coastal jet and forms cyclonic eddies~\citep{hagen1996near}, while the subduction of the filament into deeper layers highlights its role in mass and energy export~\citep{cape_ghir_sangra}.

Other capes, such as Jubi and Bojador, exhibit similar dynamics where the coastal wind angle and bathymetric irregularities locally intensify upwelling. Atmospheric forcing, topographic constraints, and vorticity adjustments collectively sustain the biological productivity and mesoscale variability unique to the CCUS~\citep{PELEGRI20053}.

\subsection{High-Resolution L4 sea surface temperature reprocessed}

SST data is invaluable for validating models as it evaluates air-sea interactions and vertical mixing while providing insights into the accuracy of model parameterizations and external forcing fields~\citep{Mourre2018AssessmentOH}. In this study, we use the L4 SST reprocessed product (SST\_ATL\_SST\_L4\_REP\_OBSERVATIONS\_010\_026) from the European Union Copernicus Marine Service (CMEMS) for the Atlantic Ocean around Iberia, Biscay, Ireland (IBI), and the northwestern European shelf domain~\citep{CopernicusMarineService2023}. It covers nearly 40 years of daily SST data collected by satellites from 1982 to 2020. This high-resolution product, at \numprint{0.05} degrees resolution ($\approx \numprint{5.55}\:km$), covers the entire IBI domain ($\approx \numprint{17442538}\:km^2$), ranging from \numprint{8.93}° to \numprint{61.98}° latitude and \numprint{-20.97}° to \numprint{12.98}° longitude. It is provided in NetCDF-4 format and is represented in a standard coordinate system (WGS 84/World Mercator). This resource is produced by Ifremer in France and is updated annually.

The L4 product is built from the L3S product SST\_ATL\_PHY\_L3S\_MY\_010\_038 using the inter-calibration method described in~\cite{piolleQUIDSSTTAC2023}. Satellite measurements of the SST come from various sources, such as NASA, NOAA, EUMETSAT OSI-SAF, and ESA. These sources are combined using this inter-calibration to create a unified dataset. Each data source includes information about the sensor-specific error statistics to help assess data quality. This information and quality flags are used to identify and select the least reliable data. 

For each day, a correction for SST values is estimated to account for discrepancies between the satellites to ensure a consistent daily dataset. For each satellite, a large-scale bias field is determined by comparing the observations of the satellite with the daily reference field. This bias is then smoothed using a Gaussian filter, which helps reduce noise and other irregularities. The smoothed bias is subtracted from the original SST values, resulting in adjusted temperatures that are more accurate and reliable. This adjustment is essential for correcting systemic errors and improving the overall quality of the data.

Once the SST values are adjusted, the single-sensor composite files, L3C, are combined into a multi-sensor composite file, L3S. This merging process ranks sensors based on their accuracy, determined through comparisons with direct measurements. Each cell in the final grid contains data from the best sensor available, ensuring the highest quality SST values.

Our study area is a subdomain within the IBI region, ranging from \numprint{19.55}°N to \numprint{34.525}°N and \numprint{20.97}°W to \numprint{5.975}°W, covering an area of approximately $\numprint{2462475}\:km^2$. This region is represented by a grid of $300 \times 300$ cells. The temporal range of the data used spans from January 1, 1982, to December 31, 2020, corresponding to a total of $\numprint{14245}$ frames (daily images) and a storage size of \numprint{10.25}\:GB. The final L4 product aims to depict a gap-free daily mean sub-skin SST field in Kelvin at a depth of 20\:cm. Therefore, there are no considerable differences between the surface and potential temperatures at this depth.

We preprocessed the dataset by handling missing values for two purposes: first, we used them to calculate a static binary land-water mask, which we smoothed using a Gaussian filter; second, we filled the missing values with the average SST value. We used the average because the model requires complete input data without gaps, and the mean provides a neutral value that avoids introducing strong gradients or artificial patterns.

\section{Forecast methods}
\label{se:forecast_methods}
This section describes the forecasting approaches used to predict SST in the CCUS region, with all models validated against satellite-derived L4 SST data (serving as ground truth). We compare our graph neural network against three baselines: two numerical ocean models and one machine learning approach. First, PSY4V3R1, the operational numerical forecast system that generates daily 10-day predictions using the NEMO platform, represents the state-of-the-art deterministic ocean forecasting. Second, GLORYS, a high-resolution global ocean reanalysis based on the same NEMO framework but enhanced through reanalyzed atmospheric forcing data. Regarding machine learning baselines, we include a ConvLSTM neural network, which combines convolutional layers with long short-term memory (LSTM)~\citep{lstm} units to capture spatiotemporal patterns in SST evolution.

Then, we explain our graph neural network in detail, a model that leverages a multiscale graph representation to improve predictions while reducing computational cost. We discuss the modifications to adapt the model for regional oceanography, including adjustments to the spatial structure and loss function optimization. The specific configurations and hyperparameters for the graph neural network and ConvLSTM models are detailed in~\autoref{ap:model_config}.

\subsection{PSY4V3R1 forecast}
Since October 2006, the Mercator Ocean PSY4V3R1 system has provided high-resolution global ocean monitoring and forecasting under CMEMS. With a $1/12^\circ$ ($\sim 9 km$) horizontal resolution and 50 vertical levels—offering fine-scale detail in the upper ocean—it captures essential oceanic processes for operational use. Built on NEMO v3.1~\citep{madec2008nemo}, it integrates high-frequency atmospheric forcing from the European Centre for Medium-Range Weather Forecasts (ECMWF)~\citep{os-14-1093-2018}. Its data assimilation scheme, which combines a reduced-order Kalman filter with 3D-VAR bias correction~\citep{brasseur2006seek}, ingests satellite altimetry, SST, sea ice concentration, and in situ T\/S profiles~\citep{os-9-57-2013}.

PSY4V3R1 introduces several key upgrades over its predecessor, PSY4V2. It corrects atmospheric forcing with satellite data, incorporates freshwater runoff from ice sheet melt, and applies a time-varying steric effect to improve sea level representation. The system refines mean dynamic topography with GOCE geoid data~\citep{https://doi.org/10.1029/2010JC006505}, enhances coastal accuracy with adaptive observational errors, and reduces deep ocean drifts by integrating WOA13v2 climatology. 

It also strengthens data reliability through improved T\/S profile quality control, optimized SSH increments, and assimilation of CMEMS OSI SAF sea ice concentrations~\citep{os-9-57-2013}. By aligning a $2.2 \; mm/yr$ global mass trend with contemporary sea-level rise estimates and deriving background error covariance from bias-corrected simulations, the system improves forecasting stability~\citep{chambers2017evaluation}.

The system generates 10-day ocean forecasts, including the runtime day itself, meaning it provides projections for nine days into the future~\citep{CMEMS-GLO-PUM-001-024}. Updated daily at 00 UTC, it uses the PSY4V3R1 model to predict 3D ocean variables (e.g., temperature, salinity, currents) and 2D variables (e.g., sea level, ice thickness, mixed layer). The atmosphere–ocean coupled NEMO model requires at least 5.5 hours to complete a 6-day forecast when utilizing 864 processors~\citep{gmd-14-1081-2021}. However, in practice, the data assimilation process used in PSY4V3R1 introduces additional computational overhead, so the total runtime for a full cycle may extend to several more hours, even when using a few hundred cores.

\subsection{GLORYS12V1 Reanalysis}
\cite{GLORYS12} describes GLORYS12 as a high-resolution global ocean reanalysis system based on the ocean and sea ice NEMO models~\citep{NEMO}, starting its simulation in 1991. The system operates NEMO on a quasi-isotropic grid with a $1/12^\circ$ horizontal resolution and 50 vertical levels. The ocean model is coupled with and forced by the ERA-Interim~\citep{ERA-Interim} atmospheric reanalysis for surface conditions. It also benefits from reanalyzed atmospheric forcing rather than analyses and forecasts, incorporates higher-quality reprocessed observations, and includes refined data assimilation procedures.

GLORYS12 applies the singular evolutive extended Kalman (SEEK)~\citep{brasseur2006seek} filter method for data assimilation, integrating various sources of information (i.e., satellite sea level anomalies (SLA)~\citep{os-12-1067-2016}, satellite SST~\citep{Ezraty}, and in situ temperature and salinity (T/S) vertical profiles~\citep{os-9-1-2013, szekely2019cora}). A 3D-VAR bias correction scheme also estimates large-scale temperature and salinity biases, improving subsurface ocean variability representation.

GLORYS12 utilizes \numprint{1296} processors and completes a 7-day simulation in approximately four hours of computational time. However, the total runtime extends to 14 days because the model employs the incremental analysis update (IAU) method~\citep{os-9-57-2013} to assimilate corrections.

\subsection{ConvLSTM}
The Convolutional LSTM (ConvLSTM) cell with peephole connections, introduced by~\cite{2015arXiv150604214S}, is a specialized recurrent neural network (RNN) that combines the capabilities of CNNs to extract spatial correlations with the gating mechanisms of peephole LSTMs to capture temporal dependencies. ConvLSTM is widely used in spatiotemporal prediction tasks, including ENSO forecasting~\citep{8851967, gmd-14-6977-2021}, nearshore water level prediction~\citep{10.3389/fmars.2024.1470320}, and tropical cyclone precipitation nowcasting~\citep{YANG2022109003}.

Researchers initially used RNNs for time-series problems but faced challenges with vanishing and exploding gradients in long sequences~\citep{279181}. LSTM networks introduced gating mechanisms to address these issues, enhancing the handling of extended sequential tasks~\citep{lstm, 818041, 861302}. ConvLSTMs further advanced this by incorporating convolutions into their gates, enabling the simultaneous capture of temporal and spatial features. 

However, modern LSTM architectures often omit peephole connections, reducing the parameter number without significantly impacting performance~\citep{7508408}. This adjustment aligns contemporary ConvLSTM implementations more closely with the standard LSTM cells. ConvLSTM replaces traditional matrix multiplications with convolutional operations in input-to-state and state-to-state transitions. This design enables the model to effectively capture spatial features by analyzing local neighboring inputs and states.

The input and forget gates regulate data flow by controlling how new data integrates with previous states to generate an updated cell state, while the output gate determines the current output. Let $\mathbf{x}^t$ be the SST map for a given time instant $t$, the ConvLSTM equations are given by:
\begin{align*}
    \mathbf{i}^t &= \sigma(\mathbf{W}_{ii} * \mathbf{x}^t + \mathbf{W}_{hi} * \mathbf{h}^{t-1} + \mathbf{b}_i), \nonumber\\
    \mathbf{f}^t &= \sigma(\mathbf{W}_{if} * \mathbf{x}^t + \mathbf{W}_{hf} * \mathbf{h}^{t-1} + \mathbf{b}_f),\nonumber\\
    \mathbf{g}^t &= \tanh(\mathbf{W}_{ig} * \mathbf{x}^t + \mathbf{W}_{hg} * \mathbf{h}^{t-1} + \mathbf{b}_g), \nonumber\\
    \mathbf{o}^t &= \sigma(\mathbf{W}_{io} * \mathbf{x}^t + \mathbf{W}_{ho} * \mathbf{h}^{t-1} + \mathbf{b}_o), \nonumber\\
    \mathbf{c}^t &= \mathbf{f}^t \odot \mathbf{c}^{t-1} + \mathbf{i}^t \odot \mathbf{g}^t, \nonumber\\
    \mathbf{h}^t &= \mathbf{o}^t \odot \tanh(\mathbf{c}^t), \nonumber \label{eq:convlstm}
\end{align*}
where $\mathbf{i}^t$ represents the input gate, which determines how much new information from the input sequence $\mathbf{x}^t$ and the previous hidden state $\mathbf{h}^{t-1}$ is allowed into the memory cell. The cell state $\mathbf{c}^t$ acts as a memory that retains relevant information over time and is updated based on the input gate and the forget gate $\mathbf{f}^t$, which controls the amount of information from the previous cell state $\mathbf{c}^{t-1}$ to retain or discard. The output gate, $\mathbf{o}^t$, determines how much information from the updated cell state $\mathbf{c}^t$ is passed to the hidden state $\mathbf{h}^t$, which serves as the final output of the cell at the current time step. The activation functions are the sigmoid, $\sigma(\cdot)$, and the hyperbolic tangent, $\tanh(\cdot)$.

Parameters $\mathbf{W}_{ii}$, $\mathbf{W}_{hi}$, $\mathbf{W}_{if}$, $\mathbf{W}_{hf}$, $\mathbf{W}_{ig}$, $\mathbf{W}_{hg}$, $\mathbf{W}_{io}$ and $\mathbf{W}_{ho}$ represent the weight matrices associated with the input, $\mathbf{x}^t$, and the hidden state, $\mathbf{h}^{t-1}$, for each gate, while $\mathbf{b}_i$, $\mathbf{b}_f$, $\mathbf{b}_g$ and $\mathbf{b}_o$ are the biases for the respective gates. The Hadamard product ($\odot$) performs element-wise multiplication, enabling selective gating at each step. While classical LSTMs rely on matrix multiplications, ConvLSTMs replace these operations with convolutions ($*$) within each gate.

\subsection{Graph neural network}
Our method is based on a GNN model for global medium-range weather forecasting, originally trained on weather reanalysis data, predicting various weather variables globally at a high resolution in under a minute. It is an adaptation of the GraphCast~\citep{lam2023graphcast} model for oceanographic forecasting. This autoregressive model predicts a new state based on two previous time steps. The processing occurs in an underlying multiscale mesh refined from an icosahedron in multiple resolutions. The neural network structure comprises an \textit{encoder}, which embeds the input-grid variables into the mesh nodes, a \textit{processor} that propagates messages through the multiscale mesh, and a \textit{decoder} that maps the forecasting back onto the grid. 

The model was originally designed for global atmospheric forecasting and employs two variable types: input and input/target. The model can predict eleven variables: five surface and six atmospheric variables, using data from the ERA5 dataset. These predictions rely on two static input variables, geopotential at the surface and land-sea mask, and five input forcing terms, including solar radiation at the top of the atmosphere and four time-related features.

We adapted this model to predict SST in a local region. This adaptation involved simplifying the model, reducing both the input and input/target variables, and relying on the four time-based features and a single static variable, the land-sea mask. Additionally, to optimize the model for regional use, we replaced the icosahedral mesh with a square curvilinear mesh that represents a curved section of a spherical surface. These modifications highlight the model’s versatility and adaptability for different scales and tasks.

The model uses graphs to simulate the relationship between SST values, represented as discrete cells in a grid. It relies on a bipartite graph $\mathcal{G =\{V^\mathrm{g}, V^\mathrm{m}, \varepsilon^\mathrm{g}, \varepsilon^\mathrm{g2m}, \varepsilon^\mathrm{m2g}\}}$, made up of two sets of nodes or subgraphs: $\mathcal{V^\mathrm{g}}$, arranged in a grid pattern, and $\mathcal{V^\mathrm{m}}$, structured in a planar and regular mesh. Only the nodes in $v^m_{i} \in \mathcal{V^\mathrm{m}}$ have bidirectional connections via edges $SS_{s,r}^m \in \mathcal{\varepsilon^\mathrm{m}}$. Additionally, these two sets of nodes are connected by edges, $SS_{s,r}^{g2m} \in \varepsilon^\mathrm{g2m}$ and $SS_{s,r}^{m2g} \in \varepsilon^\mathrm{m2g}$, modeling a directional relationship between the grid and the mesh nodes, and vice versa.

\begin{figure}[ht!]
    \centering
    \includegraphics[width=0.7\linewidth]{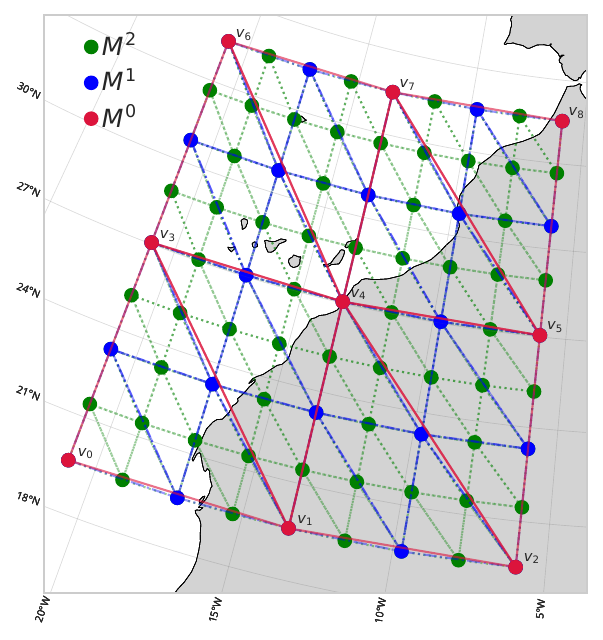}
    \caption{Representation of a multi-mesh with a refinement factor of $r = 2$, consisting of a total of 81 nodes. The nodes are grouped into three resolution levels: 9 nodes belong to $M^0$ (the coarsest scale, in red), 16 nodes to $M^1$ (the intermediate level, in blue), and 56 nodes to $M^2$ (the finest level, in green). We replaced the traditional icosahedral mesh with a curvilinear mesh based on latitude and longitude coordinates. In this approach, each node's position is explicitly described by its latitudinal and longitudinal values, ensuring a more accurate representation of the Earth's spherical geometry. New nodes are generated by refining the angular midpoint between existing nodes in spherical coordinates, avoiding distortions from planar approximations and improving spatial accuracy for regional-scale modeling.}
    \label{fig:multi_mesh}
\end{figure}

This bipartite graph defines local relationships \((\varepsilon^\mathrm{g2m}, \varepsilon^\mathrm{m2g})\) between grid node groups \(v_{i}^g \in \mathcal{V}^g\), linked through mesh nodes \(v_{i}^m\). Additionally, distant multi-scale relationships between these neighborhoods are captured by \(\varepsilon^\mathrm{m}\) connections. The number of scales, \(r\), is modeled using a multi-mesh configuration \(M^r: {\mathcal{V}^g}\), defined by embedded mesh refinements \(\{M^0, M^1, M^2, \dots, M^r\}\).~\autoref{fig:multi_mesh} shows an $M^2$ mesh with red nodes at the coarsest scale with long-range connections, blue nodes at the second scale, and green nodes at the finest scale with short-distant edges. 

The model uses a learnable algorithm called Interaction Network (IN)~\citep{battaglia2016interactionnetworkslearningobjects, 2017arXiv170601433W} to define how nodes in the graph interact with others. This IN is designed to understand relationships in complex systems~\citep{2018arXiv180601261B, 2020arXiv201003409P, 2022arXiv220207575K}. At its core, the IN relies on multilayer perceptrons (MLPs), which in the standard GraphCast implementation typically employ a latent size of 512. However, we use smaller latent sizes since our model forecasts a single oceanographic variable. This reduction in latent dimensionality offers a significant advantage, resulting in a more parsimonious model with fewer parameters. Consequently, the model becomes computationally less demanding, leading to accelerated training and inference. 

The IN mechanism facilitates sending messages from one sender node $s$ to another receiver node $r$, allowing information to flow and update the features of both the nodes and the edges across $\mathcal{G}$. The \textit{encoder}, \textit{processor}, and \textit{decoder} work through specific message-passing steps within different parts of $\mathcal{G}$. In the \textit{encoder}, nodes in $\mathcal{V}^\mathrm{g}$ send messages to nodes in $\mathcal{V}^\mathrm{m}$, $s^g \rightarrow r^m$. The \textit{decoder} then reverses this process, with nodes in $\mathcal{V}^\mathrm{m}$ sending messages back to nodes in $\mathcal{V}^\mathrm{g}$, $r^g \leftarrow s^m$. The \textit{processor} handles messages between nodes in $\mathcal{V}^\mathrm{m}$, $s^m \rightarrow r^m$, performing this step one or more times. Each additional message-passing step increases the number of model parameters, as each step requires an independent IN. 

More details about the architecture of the model are given in~\autoref{ap:GNN_model}, with the structure of the bipartite graph and the equations of the \textit{encoder}, \textit{processor}, and \textit{decoder}.

\section{Results}
\label{se:results}

\subsection{Metric-Based Evaluation and Quantitative Analysis of Forecast Skill}
The performance of the models was verified against the SST L4 satellite data and assessed through a comprehensive forecast verification framework, detailed in~\autoref{ap:scores}, based on four metrics: Root Mean Square Error (RMSE), Anomaly Correlation Coefficient (ACC), Bias, and Relative Activity (RA). These metrics were calculated following the methodology described in~\cite{RiseofDataDriven}.

\begin{figure}[ht!]
    \centering
    \includegraphics[width=1\linewidth]{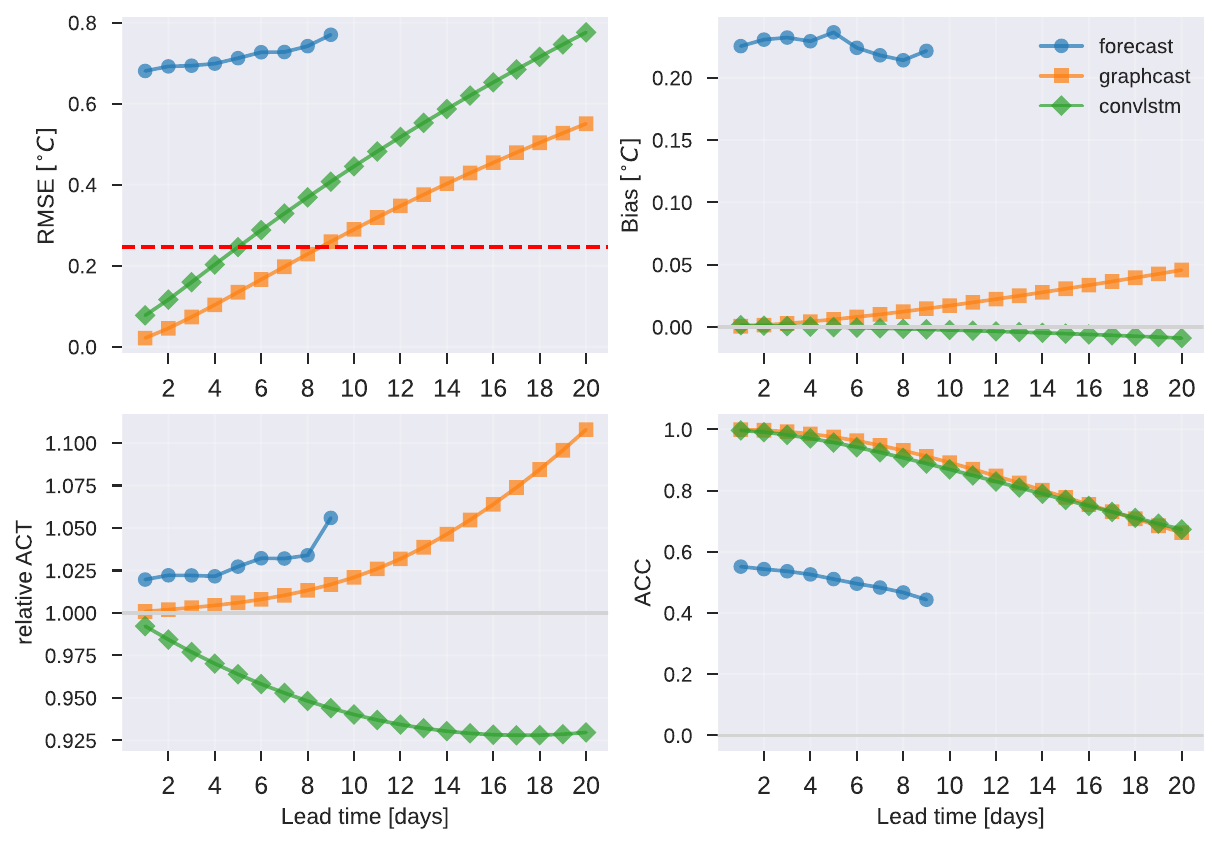}
    \caption{Performance evaluation of the models for SST, verified against satellite L4 observations. The plots display four key metrics over a 20-day forecast period: RMSE, Bias, Relative ACT, and ACC. The models compared are GraphCast (orange), ConvLSTM (green), and PSY4V3R1 (blue), while the red dashed line in the RMSE panel represents the instrumental error of the satellite data ($\pm \numprint{0.25}^\circ C$).}
    \label{fig:scores_3models}
\end{figure}

Each score was computed independently for each lead time, up to 20 days, using daily temperature fields. The evaluation was performed on a gridded domain of $300 \times 300$ points in latitude and longitude, covering the Morocco subregion. For each grid point, the scores were averaged over $365 \times 4 \times 20$ individual forecast realizations, corresponding to daily forecasts generated from 2017 to 2020, the time range of the test set, as shown in~\autoref{fig:scores_3models}. This methodology yields robust statistical estimates of model performance for each lead time.

The SST satellite product used for verification contains an instrumental error for each grid cell within the domain and for each day of the test dataset. The average of this instrumental error is $\pm \numprint{0.25}^\circ C$ across the entire dataset. This value serves as a reference threshold in the RMSE evaluation, establishing a lower bound for the expected accuracy of the predictive models. In this context, our model surpasses the instrumental error threshold on the eighth day and ConvLSTM on the fifth day, indicating differences in their ability to match the observational accuracy of the satellite product.

A quantitative comparison of the forecast Skill Score ($S_{score}$) between the two deep learning models was performed using the normalized difference of their scores. This methodology follows the approach described by~\cite{Geer01122016}. The relative RMSE (Eq. ~\ref{eq:rmse}) and Activity (Eq~\ref{eq:activity}) percentage errors were calculated as:
\begin{equation}
    S_{rmse} = \frac{\left(\overline{\text{RMSE}_B - \text{RMSE}_A}\right)}{\overline{\text{RMSE}}_B} \times 100
\end{equation}
where $A$ represents the GraphCast model, and $B$ may represents the reference GLORYS or ConvLSTM models. For the ACC (Eq.~\ref{eq:acc}) and bias (Eq.~\ref{eq:bias}) metrics, which can take negative values, the normalized difference was calculated as:
\begin{equation}
    S_{acc} = \frac{\left(\overline{\text{ACC}_A - \text{ACC}_B}\right)}{\left(\overline{1 - \text{ACC}_B}\right)} \times 100
\end{equation}

The spatial averages were computed on the $300 \times 300$ grid cells in every forecast, and the temporal averages were subsequently calculated for each day. This resulted in 20 values representing the normalized difference for each lead time. Each value represents the average of $300 \times 300 \times 365 \times 4$ data points. 

The $S_{rmse}$ and $S_{acc}$ quantify the relative performance of GraphCast compared to ConvLSTM. Those percentages indicate that our model achieves a higher or lower score than ConvLSTM at a given lead time. 

Looking at \autoref{fig:scores_3models}, we observe that, for the initial five lead times, GraphCast demonstrates a \numprint{81,7}\% to \numprint{38,4}\% higher $S_{acc}$ relative to ConvLSTM, while for the subsequent ten lead times (6 to 10), the $S_{acc}$ is 30,9\% to 7,8\% higher. In terms of $S_{rmse}$, GraphCast exhibits an error reduction between \numprint{74,6}\% and \numprint{30,0}\%  in the first ten lead times, and an improvement of \numprint{28,4}\% to \numprint{21,7}\% in the last ten lead times. These results highlight the enhanced predictive skill of the graph model across short and extended forecast horizons.

In terms of $RA$, GraphCast shows overactivity with a monotonically increasing level relative to observations, while ConvLSTM shows underactivity, with a consistently lower and decreasing level across lead times. Despite these differences, both models maintain a low mean bias across all forecast lead times. ConvLSTM’s bias remains virtually zero ($< 0,009$ °C), whereas GraphCast’s bias increase gradually up to \numprint{0,05}°C at the last lead time. In both cases, the bias stays well within the instrumental uncertainty. Note that these biases are averages over the spatial domain and all initialization dates yielding a naively results.

\subsection{Interannual model performance}
\begin{figure}[ht!]
    \centering
    \includegraphics[width=0.75\linewidth]{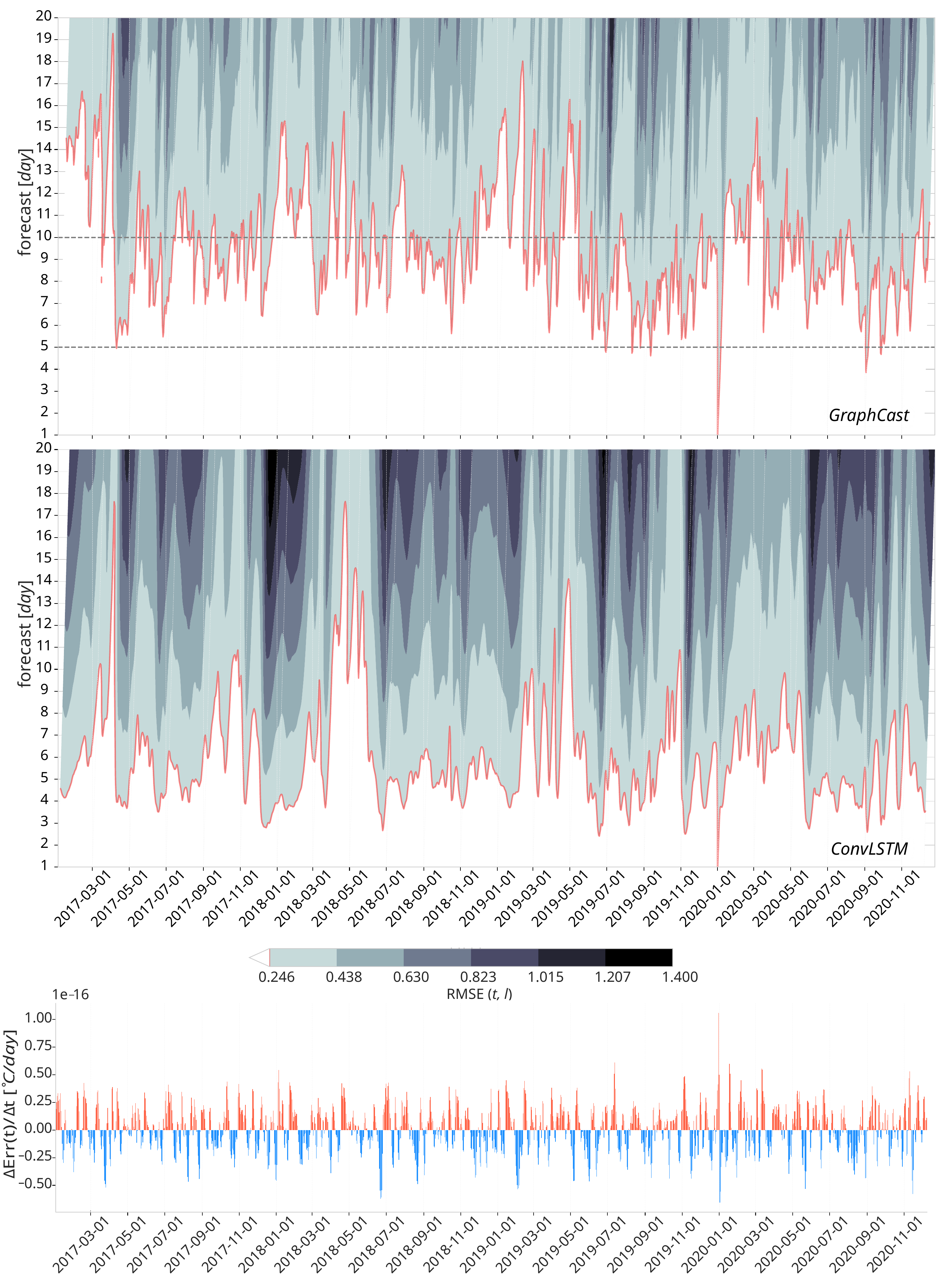}
    \caption{Predictability barrier plots of daily RMSE for SST forecasts at lead times of 1–20 days from 2017 to 2020, comparing GraphCast (top panel) and ConvLSTM (middle panel). Color shading—from pale (low RMSE) to dark (high RMSE)—represents the evolution of forecast error as a function of lead time and forecast initialization date. The solid red line indicates the mean instrumental-error threshold, while dashed grey lines mark 5- and 10-day lead-time references. All RMSE values are validated against L4 satellite SST data. The bottom panel displays time series of daily instrumental-error anomalies ($\Delta Err/\Delta t$, $^\circ C$ $day^{-1}$), with red (blue) bars indicating increasing (decreasing) errors. Notably, on 1 January 2020, both models exhibit a simultaneous RMSE spike exceeding the mean instrumental-error threshold, coinciding with a sharp positive anomaly in instrumental error—indicating a transient but substantial loss of predictability.}
    \label{fig:barrier_plot_3models}
\end{figure}

We can analyze the precision of the methods using barrier plots, where the y-axis represents the lead time, the x-axis represents the forecast day, and the color scale indicates the RMSE value. These plots depict $365 \times 4 \times 20$ predictions, where each point represents the latitude-weighted spatial average of the $300 \times 300$ grid cells for each day.~\autoref{fig:barrier_plot_3models} compares the barrier plots of the models from 2017 to 2020. 

We can use these barrier plots to assess the percentage of forecast days in which model errors exceeded the mean satellite instrumental error across different lead times and forecast realizations from January 3, 2017, to December 26, 2020. We quantified the proportion of forecast lead time days where RMSE values surpassed the instrumental error threshold (cf.~\autoref{tab:baseline_comparison}). Higher percentages indicate periods of reduced model skill, and lower values indicate improved performance.

A notable anomaly occurred on January 1, 2020, during which both models demonstrated unusually low skills. This anomaly coincided with a significant shift in satellite instrumental error, suggesting a potential link between the two events, as illustrated in~\autoref{fig:barrier_plot_3models}. The magnitude of this instrumental error shift was quantified to range between $\numprint{3.2} \sigma$ to $\numprint{5.2} \sigma$ and was compared against the time series of instrumental error variability. This comparison supports the hypothesis that observational uncertainties contributed to the degraded model performance on that day.

\begin{table}[ht!]
\centering
\caption{Percentage of forecasts exceeding the instrumental error threshold for GraphCast and ConvLSTM (2017–2020). The comparison is based on $\numprint{28719}$ RMSE values, above the instrumental error as the baseline.}
\begin{tabular}{cccc}
\hline
\textbf{Year} & \textbf{N} & \textbf{GraphCast} \% & \textbf{ConvLSTM} \% \\ 
\hline
2017 & \numprint{7070} & \numprint{51.0} & \numprint{72.0} \\ 
2018 & \numprint{7300} & \numprint{52.0} & \numprint{71.0} \\ 
2019 & \numprint{7300} & \numprint{57.0} & \numprint{73.1} \\ 
2020 & \numprint{7049} & \numprint{59.3} & \numprint{75.2} \\ 
\hline
\end{tabular}
\label{tab:baseline_comparison}
\end{table}

Both models recorded the highest percentage in 2020, with \numprint{59.3}\% of RMSE values above the threshold for GraphCast and \numprint{75.2}\% for ConvLSTM. This achieved its lowest value in 2018, at 71\%, while GraphCast recorded its lowest in 2017 at 51\%. This highlights a period of relatively better performance for GraphCast, particularly in 2017, where it was significantly better than ConvLSTM. The results underscored GraphCast's improved consistency and accuracy over ConvLSTM across the evaluated years.

GraphCast produced a four-year RMSE average of $\numprint{0.42}^\circ C$ with a standard deviation of $\numprint{0.13} ^\circ C$. The model showed a slight reduction in 2018 ($\numprint{0.40}^\circ C$) and the highest increment in 2020 ($\numprint{0.43}^\circ C$). In contrast, ConvLSTM yielded higher RMSE values, with a four-year average of $\numprint{0.53}^\circ C$ and a standard deviation of $\numprint{0.2}^\circ C$. The model exhibited its lowest RMSE in 2019 ($\numprint{0.52} ^\circ C$) and the highest value in 2018 ($\numprint{0.54} ^\circ C$).

Additionally, we calculated the skill RMSE of GraphCast compared to ConvLSTM, following the same methodology described in the previous section. The $S_{rmse}$ shows that GraphCast consistently outperformed ConvLSTM in the test period, with improvements ranging from approximately \numprint{19.4}\% to \numprint{26.5}\%, with the largest improvement in 2018. GraphCast obtained lower RMSE and more values under the instrumental error threshold, consistently indicating superior predictive performance compared to ConvLSTM in all years.

\begin{figure}[ht!]
    \centering
    \includegraphics[width=1\linewidth]{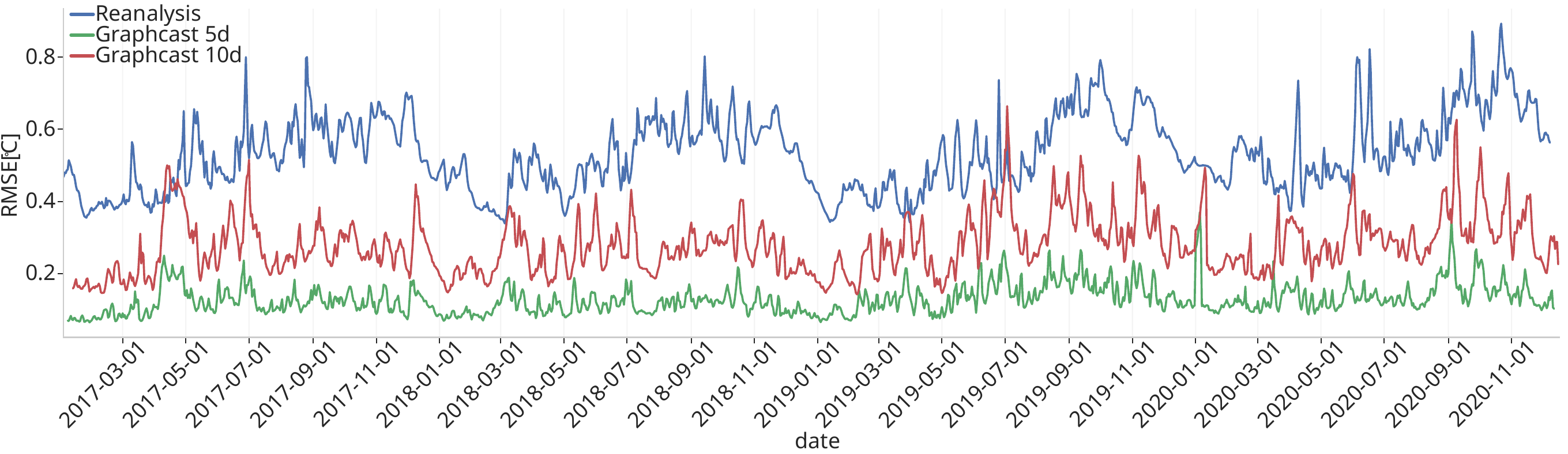}
    \caption{Time series of RMSE for SST predictions over the test period. RMSE values from GraphCast forecasts at 5-day (green) and 10-day (red) lead times are shown alongside those from the GLORYS reanalysis (blue). GraphCast consistently exhibits lower RMSE throughout the period, indicating reduced prediction error relative to the reanalysis.}
    \label{fig:time_series_gc_reanalysis}
\end{figure}

We compare the RMSE time series of GraphCast for 5-day and 10-day forecast lead times with the RMSE of the GLORYS reanalysis data in the test period, as illustrated in~\autoref{fig:time_series_gc_reanalysis}. 
The results demonstrate significant improvements in forecast skill: for the 5-day lead time, GraphCast achieves a 75,5\% reduction in RMSE compared to the reanalysis; similarly, for the 10-day lead time, GraphCast shows a 47\% lower RMSE than the reanalysis. These findings highlight GraphCast's superior performance in reducing prediction errors across the test period. Additionally, GraphCast generates a 20-day SST forecast in approximately 140 seconds on a Quadro RTX 4000 GPU with 8 GB RAM, showcasing its remarkable computational efficiency. In contrast, GLORYS, which simulates a comprehensive suite of ocean variables—including salinity, velocity components, and others—takes approximately 4 hours to complete a 7-day simulation. 

\subsection{Seasonal analysis}
\begin{figure}[ht!]
    \centering
    \includegraphics[width=0.75\linewidth]{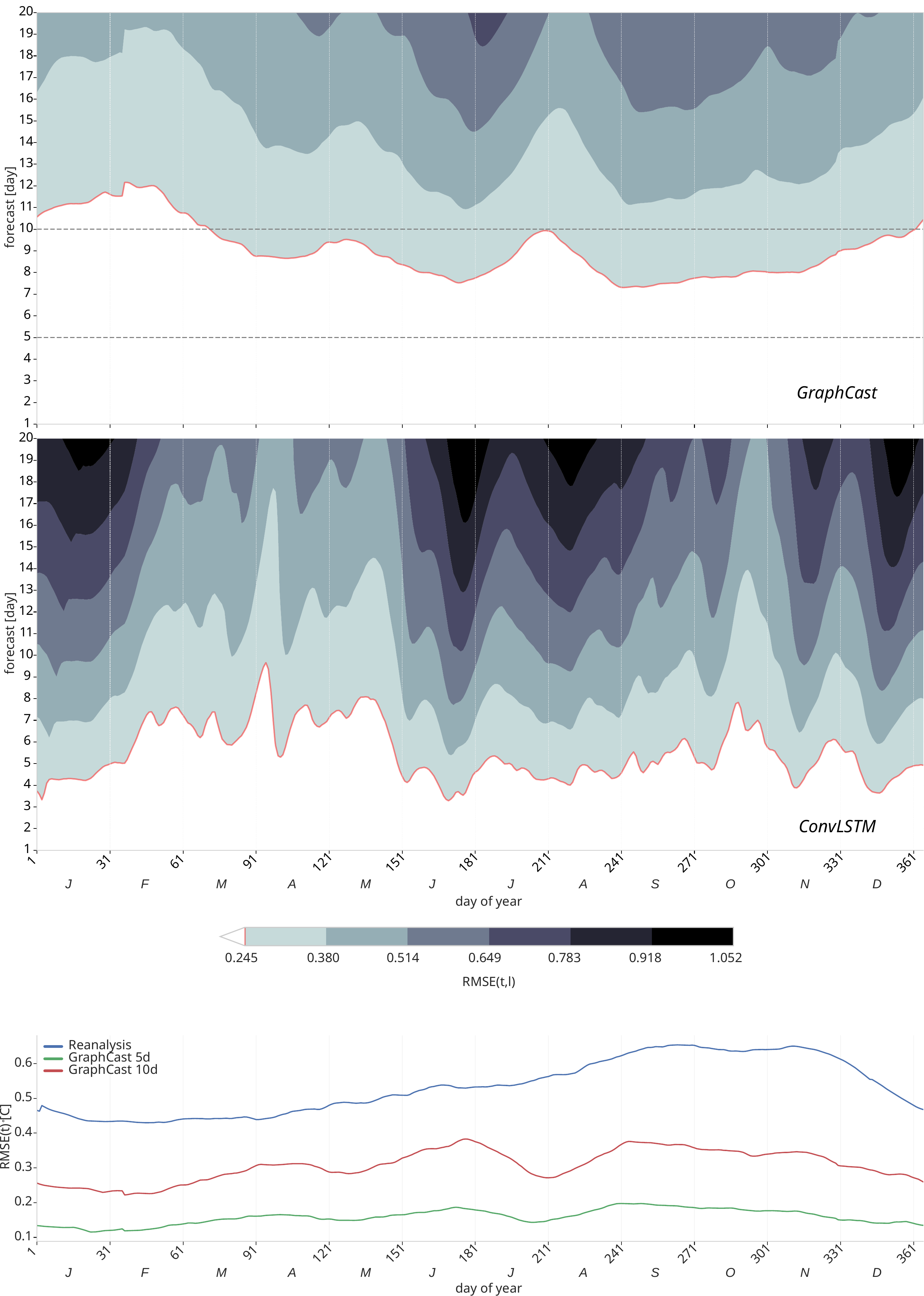}
    \caption{Seasonal predictability barrier plots showing daily RMSE for SST forecasts at lead times of 1–20 days, averaged using a 90-day moving window ($\pm 45$ days) to highlight seasonal patterns in forecast skill. The top panel corresponds to GraphCast and the middle panel to ConvLSTM. Color shading—from pale (low RMSE) to dark (high RMSE)—indicates forecast accuracy as a function of lead time and day of year. The solid red line marks the mean instrumental-error threshold, and dashed grey lines indicate 5- and 10-day lead-time references. All RMSE values are validated against Level-4 satellite SST data. The x-axis spans the full calendar year (days 1–365), with each point representing the center of a 90-day moving window, averaged across four years (2017–2020) to capture the seasonal cycle of forecast skill. The bottom panel presents the seasonal cycle of RMSE at selected lead times (5 and 10 days) for GraphCast, along with the corresponding RMSE from the reanalysis (blue), providing a reference for error magnitude across the year. These time series correspond to horizontal slices through the top panel and emphasize periods of relatively higher or lower model skill.}
    \label{fig:barrier_plot_doy_3models}
\end{figure}

To further investigate the seasonal patterns in the model, we used RMSE barrier plots based on a three-month moving window average centered on each day of the year. We apply a 90-day smoothing window to the daily spatially latitude-weighted RMSE values from the barrier plots of each model, spanning the four years of the test set. The window included 45 days before and after the target day, accumulating 90 days, repeated across the four years. Each value represents an average of $300 \times 300 \times 90 \times 4$ RMSE values, resulting in $365 \times 20$ data points per day of the year and per lead time.~\autoref{fig:barrier_plot_doy_3models} shows the seasonal barrier plots.

The seasonal performance of the models was evaluated by calculating the percentage of days within each season where the RMSE exceeded the mean instrumental error. ConvLSTM recorded the highest rate in DJF and JJA, with \numprint{19,1}\% and \numprint{20,2}\% of RMSE values above the threshold, respectively. GraphCast, on the other hand, showed its peak percentage in JJA and SON, with 15,4\% and 15,6\% of values above the threshold, respectively. GraphCast demonstrated its lowest rate in DJF, at 12\%, while ConvLSTM achieved its best performance in MAM, with 17\% of values above the threshold. These results highlight seasonal variations in model performance, with GraphCast showing relatively better consistency across seasons than ConvLSTM.

Additionally, the mean RMSE for each season was computed to assess the seasonal performance of the models (cf. ~\autoref{tab:seasonal_rmse_models}). GraphCast exhibited the highest RMSE of $\numprint{0.44}^\circ C$ with a standard deviation of $\numprint{0.1}^\circ C$ during SON, while its lowest RMSE of $\numprint{0.35}^\circ C$ with a standard deviation of $\numprint{0.07}^\circ C$ occurred in DJF. In contrast, ConvLSTM produced its highest RMSE of $\numprint{0.6}^\circ C$ with a standard deviation of $\numprint{0.2}^\circ C$ in JJA, and its lowest RMSE of $\numprint{0.4}^\circ C$ with a standard deviation of $\numprint{0.1}^\circ C$ in MAM. These findings align with the previous analysis, reinforcing that GraphCast consistently outperforms ConvLSTM across seasons, particularly in winter (DJF), where it achieves the lowest rate.

\begin{table}[ht!]
    \centering
    \caption{Seasonal RMSE for GraphCast and ConvLSTM models. The table presents the mean ($\mu$) and standard deviation ($\sigma$) of RMSE values for each season: December-January-February (DJF), March-April-May (MAM), June-July-August (JJA), and September-October-November (SON). GraphCast and ConvLSTM performance is compared across seasons, highlighting variations in prediction accuracy and consistency.}
    \begin{tabular}{lcccc}
        \hline
        \multirow{2}*{\textbf{Season}} & \multicolumn{2}{c}{\textbf{GraphCast}} & \multicolumn{2}{c}{\textbf{ConvLSTM}} \\
        \cline{2-5}
        & $\mu$ ($^\circ C$) & $\sigma$ ($^\circ C$)& $\mu$ ($^\circ C$)& $\sigma$ ($^\circ C$)\\ 
        \hline
        DJF & \numprint{0.35} & \numprint{0.07} & \numprint{0.58} & \numprint{0.20} \\
        MAM & \numprint{0.38} & \numprint{0.08} & \numprint{0.40} & \numprint{0.10} \\
        JJA & \numprint{0.43} & \numprint{0.11} & \numprint{0.60} & \numprint{0.20} \\
        SON & \numprint{0.44} & \numprint{0.10} & \numprint{0.50} & \numprint{0.15} \\
        \hline
    \end{tabular}
    \label{tab:seasonal_rmse_models}
\end{table}

Next, we calculate the relative improvement of GraphCast to ConvLSTM by estimating the $S_{rmse}$ for each season and lead time. The results demonstrate that GraphCast consistently outperformed ConvLSTM across all seasons. The most significant improvement was observed in DJF, with a \numprint{38.5}\% reduction in RMSE, while the smallest improvement occurred in MAM, with a rate of 7\%. GraphCast also achieved lower RMSE values in the remaining seasons, with \numprint{28.4}\% in JJA and \numprint{10.6}\% in SON compared to ConvLSTM. These findings align with seasonal performance trends identified earlier, further emphasizing GraphCast's superior accuracy and consistency across all seasons, particularly in winter and summer.

Additionally, the same moving window averaging procedure was applied to the RMSE of the GLORYS reanalysis dataset, verified against satellite SST, to establish a reference climatology for the reanalysis. The seasonal RMSE time series for GraphCast at 5-day and 10-day lead times was then compared to the GLORYS reference, providing a comparative view of how well these forecast lead times reproduced the seasonal cycle, as shown in~\autoref{fig:barrier_plot_doy_3models}c. The normalized difference between GraphCast and GLORYS was calculated to quantify the relative skill of GraphCast in capturing the seasonal cycle compared to the reanalysis data.

We computed the $S_{rmse}$ of GraphCast relative to the GLORYS reanalysis for each forecast lead time to evaluate model performance further. The results demonstrate significant improvements in forecast accuracy: for the 5-day lead time, GraphCast achieved the highest reduction in RMSE during DJF and SON, with a rate of \numprint{77.4}\% for both seasons, while a rate of \numprint{73.0}\% and \numprint{75.3}\% were observed for MAM and JJA, respectively. Similarly, for the 10-day lead time, GraphCast showed the highest skill in DJF and SON, at \numprint{51.2}\% and \numprint{50.7}\%, respectively, with \numprint{42.4}\% and \numprint{46.0}\% for MAM and JJA. 
This comparison highlights the seasonal variability in model performance and emphasizes the extent to which GraphCast outperforms GLORYS across different seasons and forecast horizons.

\subsection{Analysis of the spatial accuracy of models forecasts} 

\begin{figure}[ht!]
    \centering
    \includegraphics[width=1\linewidth]{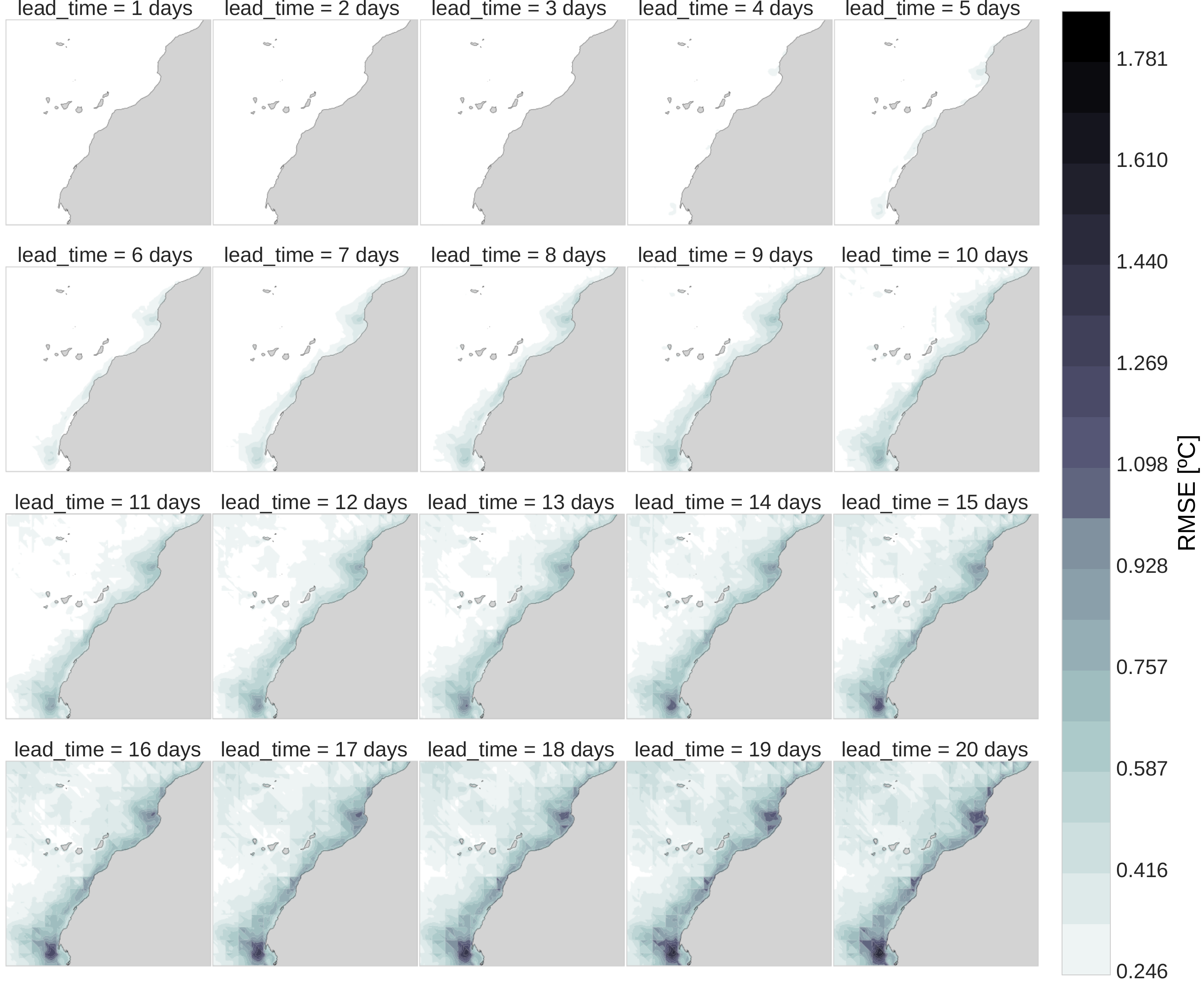}
    \caption{Point-wise RMSE average for 20 lead times forecasts of the GraphCast model. Each map represents the average RMSE at each point of our area of study for 365 days and 4 years of the test set. The images show the results of the 1 to 20 lead times from left to right and top to bottom.}
    \label{fig:rmse_location_graphcast}
\end{figure}

\begin{figure}[ht!]
    \centering
    \includegraphics[width=1\linewidth]{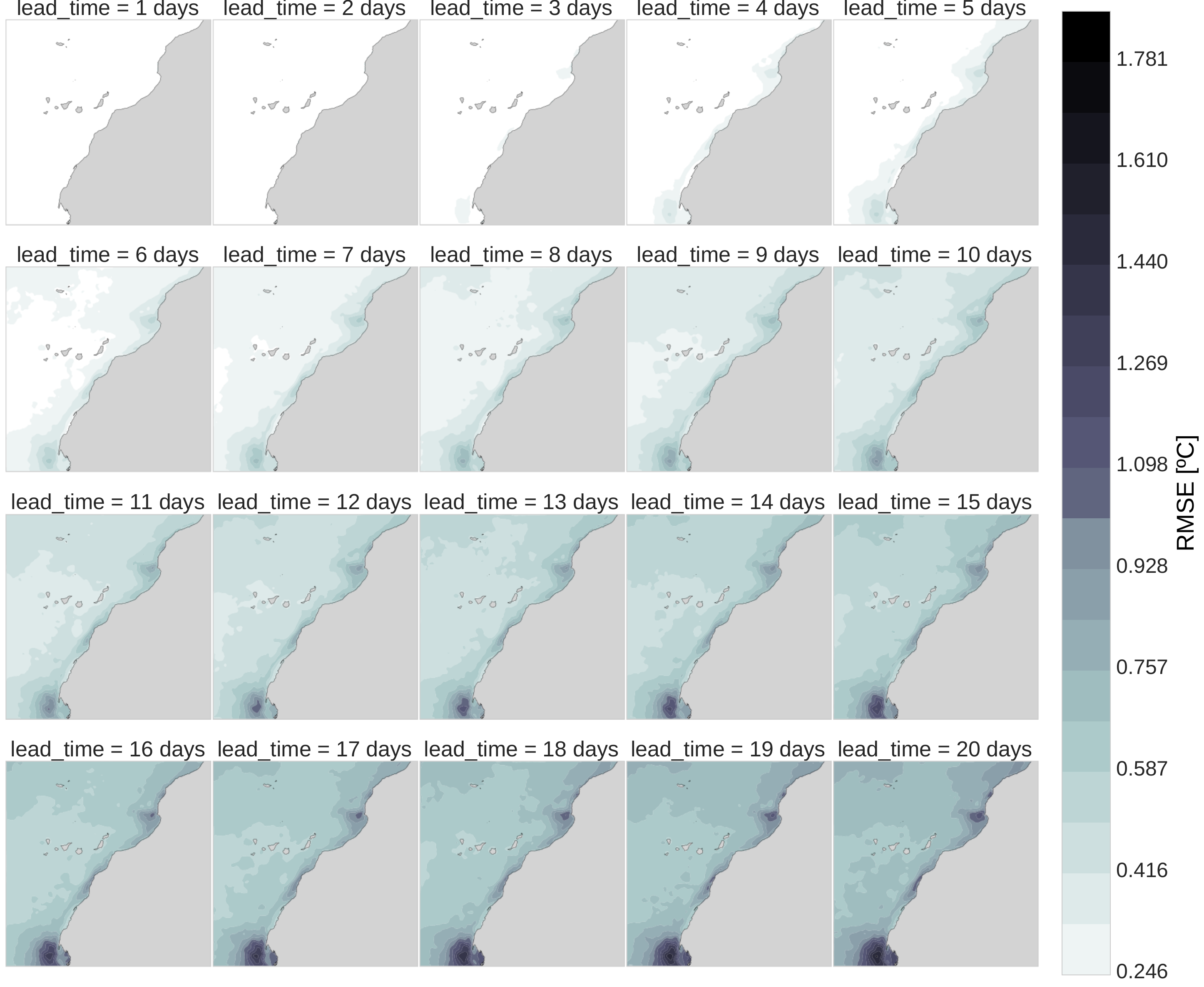}
    \caption{Point-wise RMSE average for 20 lead times forecasts of the ConvLSTM model. Each map represents the average RMSE at each point of our area of study for 365 days and 4 years of the test set. The images show the results of the 1 to 20 lead times from left to right and top to bottom.}
    \label{fig:rmse_location_convlstm}
\end{figure}

In this section, we analyze the accuracy of models in the area of study. We generated RMSE maps for each lead time, where each grid cell represents the temporal average of $365 \times 4$ RMSE values, corresponding to the four years of the test set. This averaging procedure results in 20 RMSE maps.~\autoref{fig:rmse_location_graphcast} shows the corresponding error maps for GraphCast and~\autoref{fig:rmse_location_convlstm} the maps for the ConvLSTM model.

The first step of the analysis quantifies the percentage of ocean grid cells, out of a total of $\numprint{49061}$, where the RMSE exceeded the mean instrumental error for each lead time and model. This metric provides insight into the spatial extent of regions with significant prediction errors. From the first lead time, ConvLSTM showed \numprint{0.9}\% of cells above the instrumental error threshold, while GraphCast had no cells exceeding this error. By the fourth lead time, GraphCast surpassed the instrumental error in \numprint{1.05}\% of cells, whereas ConvLSTM had a rate of \numprint{12.54}\% of cells above the threshold. Notably, ConvLSTM reached 100\% of ocean cells exceeding the instrumental error by the eighth lead time, while GraphCast remained significantly lower at \numprint{22.96}\% for the same lead time. By the twentieth lead time, GraphCast produced \numprint{99.01}\% of ocean cells exceeding the instrumental error. Therefore,  GraphCast maintains lower prediction errors across a larger spatial extent and longer lead times than ConvLSTM.

The spatial RMSE average reveals that GraphCast consistently exhibits lower values than ConvLSTM across all lead times. At the first lead time, GraphCast achieves $\mu=\numprint{0.02}^\circ C$ and  $\sigma=\numprint{0.008}^\circ C$, while ConvLSTM shows $\mu=\numprint{0.05}^\circ C$ and  $\sigma=\numprint{0.054}^\circ C$). However, the standard deviation of GraphCast grows more rapidly than that of ConvLSTM, reaching $\mu=\numprint{0.21}^\circ C$ and $\sigma=\numprint{0.094}^\circ C$ by the eighth lead time, compared to ConvLSTM's $\mu=\numprint{0.36}^\circ C$ and $\sigma=\numprint{0.088}^\circ C$). Beyond the eighth lead time, GraphCast's standard deviation remains higher than ConvLSTM, while its mean is lower. At the last lead time, GraphCast achieves a smaller mean ($\mu=\numprint{0.5}^\circ C$) than ConvLSTM ($\mu=\numprint{0.76}^\circ C$), but its standard deviation ($\sigma=\numprint{0.218}^\circ C$) is larger ($\sigma=\numprint{0.13}^\circ C$). Therefore, GraphCast maintains lower prediction errors despite its higher variability in later lead times. 

We also computed the spatial average $S_{rmse}$ relative to ConvLSTM. The mean reduction in RMSE across all lead times is \numprint{43.86}\%, with the highest reduction of \numprint{62.48}\% at the first lead time and the lowest reduction of \numprint{35.73}\% at the last lead time. This indicates that GraphCast provides consistently small errors over ConvLSTM across all forecast horizons.

The spatial analysis also allows us to identify regions with high RMSE values. Visual inspection of the maps revealed that the areas with the highest values for both models were concentrated near prominent capes, which collectively accounted for more than \numprint{18.6}\% of the total RMSE across all lead times relative to the entire domain. Specifically, Cape Ghir contributed with a rate of \numprint{6.8}\%, Cape Bojador with \numprint{4.0}\%, and Cape Blanco with \numprint{7.8}\% to the overall RMSE. Despite these challenges, GraphCast demonstrated superior performance in these regions compared to ConvLSTM. On average, GraphCast achieved significant reductions in RMSE relative to ConvLSTM, with improvements of \numprint{22.2}\% at Cape Ghir, \numprint{24.9}\% at Cape Bojador, and \numprint{28.5}\% at Cape Blanco. Therefore, GraphCast can handle complex regional dynamics better than ConVLSTM, particularly in areas with high prediction errors.

Finally, we computed the spatial $S_{rmse}$ relative to the time-average GLORYS reanalysis, with an full domain average RMSE of $\numprint{0.48} ^\circ C$. This metric quantifies the domain-averaged skill in RMSE between GraphCast and the reanalysis. For the 5-day lead time, the GraphCast average RMSE is smaller by a rate of \numprint{74,2}\% relative to the reanalysis across the entire domain. At the 10-day lead time, the error is reduced by a \numprint{44,1}\%, reflecting diminished but still significant improvements. These results demonstrate GraphCast’s ability to outperform the reanalysis benchmark, with higher performance at shorter lead times. The same analysis was applied to each coastal cape and, for the 5-day lead time, GraphCast improved the average RMSE by a \numprint{69,7}\% at Cape Ghir, \numprint{78,6}\% at Cape Bojador, and \numprint{78,5}\% at Cape Blanc relative to the reanalysis. At the 10-day lead time, the reductions were \numprint{36,8}\%, \numprint{53,5}\%, and \numprint{51,1}\%, respectively. This shows that GraphCast provides consistently better estimates than the reanalysis data. 

\section{Discussion}
\label{se:discussion}
The results presented in the previous section highlight key differences in performance between the two MLOP models—GraphCast and ConvLSTM—and the GLORYS reanalysis, along with the instrumental error baseline. GraphCast shows consistently lower RMSE and higher ACC values in all lead times, indicating superior predictive capability in the medium-range forecasts. 

Between 2017 and 2020, both deep learning models showed an increasing number of days when RMSE surpassed the instrumental error threshold, though with different patterns. GraphCast outperformed ConvLSTM overall but had a sharper rise in exceedances (from 51.0\% to 59.3\%; $+\Delta 8.3$\%) compared to ConvLSTM’s slower increase (from 72.0\% to 75.2\%; $+\Delta 3.2$\%). This suggests a notable limitation of GraphCast, especially in 2019 ($+\Delta5.0$\% interannual change): although it captures fine-scale features well, it is more sensitive to initial errors. Small inaccuracies, such as those from upwelling zones, can grow unpredictably over time, especially under strong mesoscale activity. This pattern corresponds with observed upwelling trends in the California Current Upwelling System (CCUS)~\citep{CROPPER201494}, where summer upwelling has intensified in the permanent (21–26°N) and weak permanent (26–35°N) zones.

In terms of forecast quality, ConvLSTM reduces sensitivity to upwelling-driven variability by generating smoother outputs through anomaly averaging. This behavior is evident in the RA score: ConvLSTM tends to be underactive, while GraphCast produces noisier, overactive forecasts. However, this smoothing limits ConvLSTM’s ability to resolve fine-scale dynamics, making it less suitable for high-resolution applications. By contrast, the GNN captures spatial dependencies more effectively, achieving lower RMSE and higher ACC across most lead times. Yet, GraphCast’s higher standard deviation at later lead-times reveals sensitivity to local ocean variability, while ConvLSTM’s convolutional structure inherently suppresses small-scale fluctuations. This trade-off prioritizes stability over accuracy—ConvLSTM’s visually coherent forecasts lack the granularity to capture localized dynamics, whereas GraphCast’s graph-based connectivity preserves fine-scale patterns.

Integrating GraphCast fine-scale representation with ConvLSTM stabilizing smoothing could mitigate error propagation while retaining critical dynamic structures. Their divergent performances stem from fundamental architectural differences: GraphCast graph-based approach resolves ocean patterns more precisely, while ConvLSTM convolutions blur spatially nuanced features.

GraphCast demonstrates the transformative potential of MLOP, achieving 75,5\% and 47\% RMSE reductions compared to GLORYS at 5-day and 10-day lead times while generating 20-day SST forecasts in just \numprint{2,3} minutes. In contrast, GLORYS provides a comprehensive multi-variable ocean state representation, including salinity and velocity components, but requires 4 hours for a 7-day simulation. Despite its broad scope, GLORYS incorporates uncertainties stemming from data assimilation methods, parameterizations, and model biases. GraphCast’s exceptional speed makes it a powerful tool for real-time forecasting. Expanding its capabilities to include variables like salinity and currents is crucial for broader oceanographic applications.

On January 1, 2020, both MLOP models experienced a sharp drop in forecast skill, coinciding with a significant increase in satellite instrumental error. This highlights the strong dependence of data-driven ocean forecasts on the quality of initial conditions. As noted by~\cite{RiseofDataDriven}, initializing with higher-quality data, such as operational IFS analyses instead of ERA5, can significantly improve forecast accuracy. The substantial observational errors introduced at initialization likely degraded forecast performance, depending on each model’s sensitivity to error growth, where the ocean’s chaotic nature amplifies small discrepancies. This event underscores the vulnerability of machine learning models to sporadic observational anomalies and reinforces the need for robust quality-control procedures, improved error modeling, and advanced data assimilation strategies. 

The seasonal analysis using a three-month moving window average shows that GraphCast consistently outperforms both ConvLSTM and the GLORYS reanalysis, although performance varies by season. The deep learning models exhibit their weakest performance during summer (JJA),when upwelling intensifies due to the seasonal shift of the Azores High and the Inter-Tropical Convergence Zone (ITCZ)~\citep{wooster1976seasonal, CROPPER201494}. Despite these challenging conditions, GraphCast reduces RMSE by 28.4\% compared to ConvLSTM and by 75.3\% relative to GLORYS, suggesting that its graph-based architecture more effectively captures oceanographic processes linked to intensified upwelling.

In contrast, ConvLSTM performs notably worse in JJA due to its convolutional architecture, which smooths the forecasts and filters out high-frequency variability. While this reduces noise, it also suppresses the sharp gradients and nonlinear interactions characteristic of active upwelling regimes, limiting its ability to resolve fine-scale summer dynamics. In winter (DJF), when upwelling is weaker and ocean conditions are more stable, this smoothing becomes less detrimental, and GraphCast’s advantage becomes even more pronounced, excelling at capturing steady-state patterns. During spring (MAM) and autumn (SON), both models face similar challenges, and their performance converges. 

The spatial distribution of errors reveals high RMSE values near Cape Ghir, Bojador, and Blanco—areas characterized by intense upwelling, complex bathymetry, and coastal current interactions. These conditions pose significant challenges for both numerical and deep learning models, which still struggle to fully resolve filament generation processes driven by wind forcing, coastline irregularities, and mesoscale instabilities. The absence of explicit atmospheric forcings (e.g., wind stress) and bathymetric inputs in the deep learning framework likely limits its ability to disentangle these mechanisms. Additionally, the resolution of satellite-derived L4 data used for training cannot capture submesoscale features ($< 10 km$), potentially introducing systematic biases. This aligns with findings from recent studies showing that global climate models underestimate upwelling intensity due to coarse spatial resolution~\citep{Bindoff2019, docquier2019impact}, highlighting the need for regional downscaling and high-resolution ocean-atmosphere coupled models~\citep{roberts2018benefits}.

A drawback of GraphCast is the emergence of triangular artifacts at longer lead times, visible as mesh-like patterns superimposed on SST predictions. Visual analysis links these artifacts to the decoder’s mesh structure, where each triangular element updates SST values within its influence area. Since neighboring triangles share only one edge and two nodes, inconsistencies arise—especially in dynamic regions like coastal upwelling zones—where adjacent elements may experience different physical processes. This highlights the difficulty of ensuring coherence across mesh elements under heterogeneous ocean conditions.

Mitigating these artifacts presents both computational and architectural challenges. Refining the mesh can improve spatial consistency by reducing each triangle’s influence area, as seen in higher-resolution graph-based models~\citep{lam2023graphcast,oskarsson2023graphbased,Seacast}, but at significant computational cost. Alternatively, increasing node connectivity in the decoder, drawing information from more than three nodes using attention mechanisms, could smooth transitions across triangles mitigating artifacts by distributing updates across a broader mesh neighborhood. A third strategy involves integrating convolutional layers into the decoder to blend localized element outputs, leveraging their proven spatial smoothing effect. This approach, supported by ConvLSTM’s artifact-free outputs and lower variability, has been effective in reducing grid artifacts in transformer-based models~\citep{archesweather}. Future work should explore hybrid architectures that combine adaptive mesh connectivity with convolutional smoothing, preserving GraphCast’s ability to resolve large-scale dynamics while enhancing robustness to localized errors.

GraphCast outperforms ConvLSTM and GLORYS in coastal upwelling zones, reducing errors by up to 28.5\% in upwelling regions, highlighting its superior ability to capture atmosphere–ocean interactions and fine-scale spatial gradients. However, persistent errors in these areas show that neither numerical nor deep learning models achieve optimal performance without targeted improvements. Despite these challenges, GraphCast's strong SST forecasting skill demonstrates the potential of machine learning for ocean prediction. Advancing toward operational viability will require reducing sensitivity to initial condition errors—especially under strong mesoscale activity—through improved data assimilation or hybrid modeling strategies. Architectural enhancements such as convolutional layers or attention mechanisms may help mitigate artifacts efficiently, while scaling to multivariate forecasts will demand balancing complexity and speed, possibly via latent representations and physics-informed regularization. Addressing regional and seasonal variability will also require adaptive graph structures and real-time forcing assimilation, with focused validation in key regions like upwelling systems to ensure robustness and reliability.

\section{Conclusion}
\label{se:conclusion}
This work presented a detailed evaluation of deep learning architectures for oceanographic forecasting, focusing on an adapted version of GraphCast applied to the Canary Upwelling System. GraphCast achieved high spatial resolution and forecasting accuracy, surpassing ConvLSTM and traditional reanalysis products like GLORYS. These results highlighted the capacity of graph-based models to resolve mesoscale features such as filaments and eddies below 20 km in scale, offering a promising direction for data-driven approaches in operational oceanography.

Nonetheless, our findings also revealed important trade-offs. While GraphCast excelled in spatial precision, it exhibited increased sensitivity to initial condition errors, especially during summer upwelling seasons when ocean variability is highest. In contrast, ConvLSTM provided greater temporal stability due to its convolutional smoothing, but at the cost of lower spatial fidelity. These differences underscore the potential of hybrid architectures that combine the strengths of both models to balance robustness and detail in multi-day forecasts.

The analysis further showed distinct seasonal and regional error patterns. Both models performed best under winter steady-state conditions and struggled during high-variability summer periods. Spatial error concentrations were particularly pronounced near coastal capes, where bathymetric complexity and submesoscale processes below the resolution of current L4 SST training data introduced persistent inaccuracies. Additionally, the dependence on SST limited the capacity of models to provide a full representation of the ocean state, highlighting the need for physics-informed, multivariate model extensions.

Future developments should focus on hybrid architectures, improved data assimilation methods, and curated high-resolution training datasets capable of capturing small-scale dynamics. Expanding the prediction scope beyond SST—incorporating currents, salinity, and other variables—will enhance model relevance for marine resource management. Regionally adaptive graph designs that account for bathymetric and dynamical physical gradients will also be essential for mitigating systematic errors in critical areas such as coastal capes. 

This study supports the growing role of machine learning in ocean prediction, while pointing out key innovations to fully realize its potential in operational settings. Our findings position data-driven ocean prediction models as scalable alternatives to traditional numerical systems, with the ability to enhance the resolution, speed, and adaptability of next-generation ocean forecasting systems.

\section*{CRediT authorship contribution statement}
\textbf{Giovanny Alejandro C-Londoño:} Writing – original draft, Writing – review \& editing, Conceptualization, Formal analysis, Methodology, Investigation, Data curation, Software, Validation, Visualization. \textbf{Javier Sánchez:} Writing – original draft, Writing – review \& editing, Project administration, Supervision, Conceptualization, Formal analysis, Methodology, Resources. \textbf{Ángel Rodríguez-Santana:} Writing – review \& editing, Supervision, Conceptualization. 

\section*{Declaration of competing interest}
The authors declare that they have no known competing financial interests or personal relationships that could have appeared to influence the work reported in this paper.

%

%
\section*{Acknowledgements}
The authors would like to express their gratitude to Javier J. Sánchez-Medina from Centro de Innovación para la Sociedad de la Información (CICEI) at the Universidad de Las Palmas de Gran Canaria for providing access to their computing servers, which were essential for training the models. We also extend our sincere thanks to Mercator Ocean International (MOi) for sharing the PSY4V3R1 model forecasting data, which was provided upon request. Their support and collaboration were invaluable to the success of this research.

\section*{Declaration of Generative AI and AI-assisted technologies in the writing process}
The authors used ChatGPT, Gemini, and Deepseek to improve the readability and language of the manuscript during the preparation of this work. After using these tools, the authors reviewed and edited the content as needed and took full responsibility for the content of the published article.

\section*{Data availability}
The data and code used in this paper are available at the following URLs:
\begin{itemize}
    \item The European North West Shelf/Iberia Biscay Irish Seas - High Resolution L4 Sea Surface Temperature Reprocessed is available at \url{https://doi.org/10.48670/moi-00153}
    \item GLORYS is available at \url{https://doi.org/10.48670/moi-00021}.
    \item The source code of our models is publicly available at \url{https://github.com/gacuervol/regional-graphcast-sst.git} under the MIT licence.
\end{itemize}

\section*{Computational library}
The methodology is implemented in Python using the JAX library for efficient multi-GPU deep learning training. Atmospheric and oceanographic datasets are processed using Xarray, enabling scalable handling of multi-dimensional arrays, while evaluation metrics are computed with NumPy. The deep learning models are constructed using Google DeepMind Flax and Haiku frameworks, which are optimized for JAX, enabling high-performance model development.

%
\appendix
\section{Model Configuration, training protocols and experimental setup}
\label{ap:model_config}
\subsection{Training details}
The dataset was split into training, validation, and test sets using an 80/10/10 ratio, based on years rather than individual days. This temporal division preserves the representativeness of seasonal cycles in each subset. The training set covers 1982 to 2012 (31 years), the validation set spans from 2013 to 2016 (4 years), and the test set includes 2017 to 2020 (4 years). This scheme prioritizes training with historical data while testing is conducted with the most recent information.

The samples in the training set are organized into windows of three consecutive dates, with each window overlapping the next one by a single day. This approach maintains continuity in prediction trajectories while maximizing the use of available data.

We used each set to generate time forcings, providing the model with temporal context for the current day and year. As in~\cite{lam2023graphcast}, we calculated these forcings as seconds starting at the UNIX epoch from January 1, 1970, and encoded them using sine and cosine transformations to capture cyclical patterns.

For training, we organized the dataset into batches of eight samples. Each batch consists of three components: i) the input data, which includes the initial conditions for SST over two days; ii) the target data, representing the SST values for the lead times (days to predict); and iii) the forcing data, which includes the time-generated forcings for the lead times and the static land-sea mask. To ensure a stable optimization process, we applied shuffling exclusively to the training set, while the validation and test sets remained unshuffled to maintain their temporal structure.

We computed normalization factors, such as the mean, standard deviation, and standard deviation of temporal changes, using the training set. These factors were then applied to normalize the dataset, ensuring consistent scaling across all samples. This preprocessing step improved model stability and enhanced training performance by standardizing the input distribution.

We adopted the training methodologies from~\cite{lam2023graphcast}, excluding the initial warm-up phase because our experiments did not indicate a significant impact. Our training strategy comprised a 150-epoch training phase, completed in approximately 64 hours. We adjusted the learning rate using a half-cosine decay function, gradually reducing it from $10^{-3}$ to zero and updating it after each iteration. The training phase was oriented to forecasting one lead time with samples of three-time instants, $(\mathbf{x}^{t-1}, \mathbf{x}^t) \rightarrow \mathbf{x}^{t+1}$.

We used gradient descent to optimize the loss function. For adaptive moment estimation and weight decay management, we utilized the AdamW optimizer with parameters $\beta_1=\numprint{0.9}$, $\beta_2=\numprint{0.95}$, $\epsilon=10^{-8}$, and $\lambda=\numprint{0.1}$. Furthermore, we implemented gradient clipping with a maximum norm value of 32 to ensure stable training dynamics.

\subsection{Model configuration}
The ConvLSTM model employed in this study was specifically designed for spatiotemporal ocean prediction. To ensure a fair comparison with GraphCast, the model was configured and trained under similar configurations. It comprised two stacked ConvLSTM layers: the first layer contained 8 feature channels, and the second layer reduced this to 1 feature channel. Both layers utilized a $3 \times 3$ kernel with a stride of 1 and padding of 1. The input sequences had a shape corresponding to a batch size of 8, two time steps, and spatial dimensions of $256 \times 256$ with a single-channel input. The model was optimized using the AdamW optimizer, with a learning rate of $10^{-2}$ and a weight decay of \numprint{0.1}. Training was conducted over 150 epochs, with updates for single-time-step predictions.

The architecture of the graph-based approach employed three levels of multi-mesh refinement, $M^3$, implemented by recursively subdividing triangular mesh elements into smaller components. The grid structure operated at a resolution of $0,05^\circ$ derived from the satellite L4 product resolution in the training data. Spatial connectivity was regulated by a parameter that defined the neighborhood radius for each mesh node, set at $0.6$ times the edge length. Within this range, grid nodes were connected to the corresponding mesh node in the encoder, ensuring localized information aggregation. The processor module performed $6$ message-passing steps within an 8-dimensional MLP latent space, with all MLPs containing a single hidden layer. In the decoder, edge normalization for mesh-to-grid transitions was performed using the maximum edge length. This configuration effectively balanced model complexity and computational efficiency.

\subsection{Spatially-weighted Loss function}
Most weather prediction studies use the Mean Squared Error (MSE) to optimize the neural network parameters during training. Since the area of a spherical grid cell decreases toward the geographic poles, MSE is often adjusted by latitude to account for this distortion. Common approaches include weighting by the cosine of the latitude~\citep{rasp2020weatherbench, lam2023graphcast} or by using the difference between the sines of the cell's edges~\citep{rasp2024weatherbench}, both of which serve as relative indicators of grid cell area. Without such adjustments, large errors in small high-latitude cells are treated the same as small errors in much larger equatorial cells, leading to spatial imbalance in the loss. Therefore, many global prediction studies use latitude-weighted MSE. 

However, in this study, we adopt a different strategy that focuses the learning process exclusively on the ocean by applying a spatial mask derived from the structure of the data. This spatial mask assigns a weight of zero to grid cells over land and a weight of one to cells over the ocean. These binary weights are used to exclude land areas from contributing to the loss computation, effectively masking out irrelevant regions and ensuring that the loss is computed only over ocean areas. This approach is particularly well suited to our setup, which targets a geographically limited domain. In such cases, the variation in latitude across the domain is relatively small, and the differences in grid cell area are minimal. Therefore, the benefit of applying latitude-based weighting is negligible, and the masking strategy offers a more direct way to focus the training process on the regions of interest.

The loss function is thus defined as:

\begin{equation*}
\mathcal{L}_{\text{MSE}} = \frac{1}{N_{\text{Rollout}}} \sum\limits_{t = t_0 + 1}^{t_0 + N_{\text{Rollout}}} \frac{1}{|\mathbb{G}_{\text{ocean}}|} \sum\limits_{v_i \in \mathbb{G}_{\text{ocean}}} \left( \hat{x}^t_i - x^t_i \right)^2,
\end{equation*}
where
\begin{itemize}
\item $t_0$ is the initial forecast time,
\item $N_{\text{Rollout}}$ is the number of rollout steps used during training,
\item $v_i$ is a grid point defined by its latitude ($\phi$) and longitude ($\lambda$),
\item $\mathbb{G}$ is the full spatial grid, and $|\mathbb{G}| = |\phi| \cdot |\lambda|$,
\item $\mathbb{G}_{\text{ocean}} \subset \mathbb{G}$ is the subset of grid points located over the ocean, determined using a binary land–ocean mask $m_i \in \{0, 1\}$, where $m_i = 1$ indicates ocean and $m_i = 0$ indicates land.
\end{itemize}

\subsection{Experimental setup}
We used two computers to conduct our experiments. The first machine, a local workstation equipped with a single Quadro RTX 4000 GPU with 8 GB of memory, was used for hyperparameter tuning. The second machine, a remote server with 8 Quadro RTX 4000 GPUs and 8 GB of memory, was utilized for training the final model with the full dataset. The training phase on the remote server took approximately 64 hours. Our software stack included JAX, Haiku, Jraph, Optax, Jaxline, and xarray~\citep{hoyer2017xarray} for model customization and training.

During the development phase on the local machine, we conducted a hyperparameter search to explore various configurations. Specifically, we investigated the number of message-passing steps $\left\{ 2,6,7 \right\}$, latent size $\left\{ 2,4,8,16 \right\}$, and the mesh refinement level $\left\{ 2,4,6 \right\}$ denoted as $M^r$. We tested batch sizes of 8 and 16, as preliminary results indicated that batch size had a negligible impact on the model’s performance. To efficiently cover a wide range of configurations, we performed a grid search over these three hyperparameters, limiting the number of combinations while ensuring a comprehensive exploration.

Once the hyperparameters were determined, we proceeded to train the final model on the remote server using the full dataset. To optimize the training process, we parallelized it using the Distributed Data-Parallel (DDP) strategy~\citep{2020arXiv200615704L}. In this setup, each of the 8 GPUs held its copy of the model and processed a separate batch of data. After processing, we used the all-reduce operation to combine the gradients from all GPUs into a unified gradient, which was then used to update the model weights.

To accommodate the parallel processing, we added a new dimension to the data called \textit{device}, with a size equal to 8 GPUs. This modified the data dimensions to 8 devices, 8 batches, 1 to $n$ lead-time predictions, and $300 \times 300$ latitude-longitude size. Each GPU was assigned identical initialized weights, and during each training step, a forward pass was performed to calculate the loss and gradients for its assigned batch. At the end of the step, the gradients were summed on the CPU, and the resulting unified gradient was used to update the model weights across all GPUs.

This process was repeated iteratively until only residual batches remained. Since the number of residual batches was smaller than 8 (the number of GPUs), they could not be parallelized and were instead processed sequentially on the CPU.

\section{Graph Neural Network Model}
\label{ap:GNN_model}
Let $\mathbf{x}: \mathbb{R}^3\rightarrow \mathbb{R}$ be a spatiotemporal variable containing the SST volume for our period of study. Each sequence value is represented as $x_{i}^t$, with $i$ standing for node in a grid $v_i \in \mathcal{V}^g$, and $t$ the time instant. For simplicity, let $\mathbf{x}^t$ be the SST bidimensional array for time $t$ and $\mathbf{x}_{i}$ the temporal evolution of the SST in node $i$. Each node $v_i$  has a latitude-longitude position given by $(\phi_i, \lambda_i)$.

Our method is based on the GraphCast model~\citep{lam2023graphcast}, which is defined as an autoregressive model:
\begin{equation}
    \mathbf{\hat{x}}^{t+1} = f(\mathbf{x}^t, \mathbf{x}^{t-1}), 
\end{equation}
where the $\mathbf{\hat{x}}^{t+1}$ map is a forecast obtained from two previous time instants. This output can be iteratively fed into the model to predict consecutive forecastings in a roll-out manner. 

The graph of the model, $\mathcal{G}(\mathcal{V}^g, \mathcal{V}^m, \mathcal{E}^m, \mathcal{E}^{g2m}, \mathcal{E}^{m2g})$, is composed of grid nodes, $\mathcal{V}^g$, mesh nodes, $\mathcal{V}^m$, bidirectional edges connecting mesh nodes, $\mathcal{E}^m$, and directed edges from grid to mesh nodes, $\mathcal{E}^{g2m}$, and vice-versa, $\mathcal{E}^{m2g}$.

Each node in the grid, $\mathbf{v}_{i}^g \in \mathcal{V}^g$, is built from the SST values in times $t$ and $t-1$, several external forcings in times $t-1$, $t$ and $t+1$, and constant values, i.e., $\mathbf{v}_{i}^g=[x_{i}^t, x_{i}^{t-1}, \mathbf{f}_{i}^{t-1}, \mathbf{f}_{i}^{t}, \mathbf{f}_{i}^{t+1}, \mathbf{c}_{i}]$, in a given latitude-longitude position. The forcings are defined by $\mathbf{f}=[\sin(h), \cos(h), \sin(y), \cos(y)]$, with $h$ the local time of day and $y$ the year progress, normalized to $[0, 1)$. The constants are defined as
$\mathbf{c}_{i} = [m^{0/1}_{i}, \cos(\phi_{i}), \sin(\lambda_{i}), \cos(\lambda_{i})]$, with $\mathbf{m}^{0/1}$ a binary land-sea mask and no physical forcings.

Each node in the mesh, $\mathbf{v}_{i}^m \in \mathcal{V}^m$, is defined as $\mathbf{v}_{i}^m = [\cos(\phi_{i}), \sin(\lambda_{i}), \cos(\lambda_{i})]$ and the mesh edges, $\mathbf{e}^m_{s,r} \in \mathcal{E}^m$, from the sender node $s$ to the receiver node $r$, contains the following features: $\mathbf{e}^m_{s,r} = [\text{distance}(s,r), \mathbf{p}_s-\mathbf{p}_r]$, i.e. the edge length and the difference between the 3D spatial locations of nodes $s$ and $r$.~\autoref{fig:multi_mesh} shows a representation of the multi-mesh with three levels of resolution. 

Unidirectional edges from the grid to the mesh are similarly defined by $\mathbf{e}^{g2m}_{s,r} = [\text{distance}(s,r), \mathbf{p}_s-\mathbf{p}_r]$. In this case, the sender node is on the grid and the receiver is on the mesh. An edge is added between two nodes if $\text{distance}(s,r)$ is smaller than or equal to \numprint{0.6} times the length of the edges at the finest scale.

Similarly, unidirectional edges from the mesh to the grid are defined by $\mathbf{e}^{m2g}_{s,r} = [\text{distance}(s,r), \mathbf{p}_s-\mathbf{p}_r]$. An edge is created with the three mesh nodes of the triangular face that contains the grid node.

\subsection{Encoder}
The \textit{encoder} maps the input data from the latitude-longitude grid into the multi-scale mesh. It relies on multi-layer perceptrons (MLP) to embed the graph variables, $\mathbf{v}_{i}^g$, $\mathbf{v}_{i}^m$, $\mathbf{e}^m_{s,r}$, $\mathbf{e}^{g2m}_{s,r}$ and $\mathbf{e}^{m2g}_{s,r}$, into a latent space:
\begin{align}
    \mathbf{\tilde{v}}_{i}^g & = \text{MLP}_{\mathcal{V}^g}(\mathbf{v}_{i}^g),\nonumber \\
    \mathbf{\tilde{v}}_{i}^m & = \text{MLP}_{\mathcal{V}^m}(\mathbf{v}_{i}^m),\nonumber \\
    \mathbf{\tilde{e}}_{s,r}^m & = \text{MLP}_{\mathcal{E}^m}(\mathbf{e}_{s,r}^m),\nonumber \\
    \mathbf{\tilde{e}}_{s,r}^{g2m} & = \text{MLP}_{\mathcal{E}^{g2m}}(\mathbf{e}_{s,r}^{g2m}),\nonumber \\
    \mathbf{\tilde{e}}_{s,r}^{m2g} & = \text{MLP}_{\mathcal{E}^{m2g}}(\mathbf{e}_{s,r}^{m2g}).
    \label{eq:encoder_embedding}
\end{align}

Then, the information from the grid is transferred to the mesh using interaction networks (IN)~\citep{battaglia2016interactionnetworkslearningobjects}. First, the edges are updated with information from the sending and receiving nodes, 
\begin{equation}
      \mathbf{d\tilde{e}}_{s,r}^{g2m} = \text{MLP}_{\mathcal{E}^{g2m}}([\mathbf{\tilde{e}}_{s,r}^{g2m}, \mathbf{s}^g, \mathbf{r}^m]),
\end{equation}
and the mesh nodes are updated by aggregating the edges as
\begin{equation}
      \mathbf{d\tilde{v}}_{i}^{m} = \text{MLP}_{\mathcal{V}^{m}}\left([\mathbf{\tilde{v}}_{i}^m,\sum\limits_{s\in\mathcal{V}^{g}; r= \mathbf{\tilde{v}}_{i}^m}\mathbf{\tilde{e}}_{s,r}^{g2m}]\right),
\end{equation}
with $s$ the set of nodes in the grid that have an edge towards $\mathbf{v}_{i}^m$. The grid nodes are also updated as
\begin{equation}
      \mathbf{d\tilde{v}}_{i}^{g} = \text{MLP}_{\mathcal{V}^{g}}(\mathbf{\tilde{v}}_{i}^g),
\end{equation}

All MLPs in these equations are independent and do not share their parameters. Finally, the latent variables are updated with residual connections as
\begin{align}
    \mathbf{\tilde{v}}_{i}^g & \leftarrow \mathbf{\tilde{v}}_{i}^g+\mathbf{d\tilde{v}}_{i}^g,\nonumber \\
    \mathbf{\tilde{v}}_{i}^m & \leftarrow \mathbf{\tilde{v}}_{i}^m+\mathbf{d\tilde{v}}_{i}^m,\nonumber \\
    \mathbf{\tilde{e}}_{s,r}^{g2m} & \leftarrow \mathbf{\tilde{e}}_{s,r}^{g2m}+\mathbf{d\tilde{e}}_{s,r}^{g2m}.
    \label{eq:encoder_residual_update}
\end{align}

\subsection{Processor}
The \textit{processor} performs learned message-passing through several layers on the multi-mesh. First, the edges are updated through an MLP, concatenating the mesh edge with the receiver and sender nodes' latent variables as
\begin{equation}
      \mathbf{d\tilde{e}}_{s,r}^{m} = \text{MLP}_{\mathcal{E}^{m}}([\mathbf{\tilde{e}}_{s,r}^{m},\mathbf{s}^m, \mathbf{r}^m]).
\end{equation}

Mesh nodes are updated by aggregating the edges as
\begin{equation}
      \mathbf{d\tilde{v}}_{i}^{m} = \text{MLP}_{\mathcal{V}^{m}} \left(([\mathbf{\tilde{v}}_{i}^m,\sum\limits_{s\in\mathcal{V}^{m}; r= \mathbf{\tilde{v}}_{i}^m}\mathbf{\tilde{e}}_{s,r}^{m}] \right),
\end{equation}
and the variables are updated as a residual connection as
\begin{align}
    \mathbf{\tilde{v}}_{i}^m & \leftarrow \mathbf{\tilde{v}}_{i}^m+\mathbf{d\tilde{v}}_{i}^m,\nonumber \\
    \mathbf{\tilde{e}}_{s,r}^{m} & \leftarrow \mathbf{\tilde{e}}_{s,r}^{m}+\mathbf{d\tilde{e}}_{s,r}^{m}.
    \label{eq:processor_residual_update}
\end{align}

This process represents one layer of the processor. It contains multiple iterative layers with independent MLP parameters that perform several message-passing operations. 

\subsection{Decoder}
The \textit{decoder} performs the inverse mapping from the mesh to the latitude-longitude grid. First, the edges from the mesh to the grid are updated as
\begin{equation}
      \mathbf{d\tilde{e}}_{s,r}^{m2g} = \text{MLP}_{\mathcal{E}^{m2g}} [\mathbf{\tilde{e}}_{s,r}^{m2g},\mathbf{s}^m, \mathbf{r}^g] ).
\end{equation}

The grid nodes are then updated, aggregating the information of the three edges that arrive at the grid node:
\begin{equation}
      \mathbf{d\tilde{v}}_{i}^{g} = \text{MLP}_{\mathcal{V}^{g}} \left([\mathbf{\tilde{v}}_{i}^g,\sum\limits_{s\in\mathcal{V}^{g}; r= \mathbf{\tilde{v}}_{i}^g}\mathbf{\tilde{de}}_{s,r}^{m2g}] \right).
\end{equation}

A residual connection is used to update the information of the grid nodes coming from the embedding in the encoder:
\begin{equation}
        \mathbf{\tilde{v}}_{i}^g  \leftarrow \mathbf{\tilde{v}}_{i}^g+\mathbf{d\tilde{v}}_{i}^g,
    \label{eq:decoder_residual_update}
\end{equation}
and the output prediction is obtained with another MLP as
\begin{equation}
    \hat{\mathbf{y}}^{t} = \text{MLP}_{\mathcal{V}^{g}}(\mathbf{\tilde{v}}_{i}^g).
\end{equation}

Finally, the forecasting is obtained as a residual connection with the input data as
\[\mathbf{\hat{x}}^{t+1} = \mathbf{x}^{t}+\mathbf{\hat{y}}^{t}\]

All input variables are normalized to zero mean and unit variance. The output variable $\hat{\mathbf{y}}^{t}$ is multiplied by the average standard deviation of the temporal change $\mathbf{d\hat{x}}= \mathbf{\hat{x}}^{t+1}-\mathbf{\hat{x}}^{t}$ computed from the train set.

All MLPs have one hidden layer with a latent dimension of 8, the same as the output layer. The size of the output layer in the last MLP, corresponding to the \textit{decoder}, is one for the prediction of the value of the SST at each node.  

\section{Description of the evaluation metrics}
\label{ap:scores}
In this study, we used four metrics to assess the performance of the ocean prediction models: the Anomaly Correlation Coefficient (ACC), the Root Mean Square Error (RMSE), the Activity (ACT), and the Bias. While there are several ways to define those metrics, we use the approach outlined by~\cite{RiseofDataDriven} as it aligns with the scores set by the World Meteorological Organization (WMO) and the ECMWF. Below is a detailed description of each metric, including its mathematical formulation and interpretation. 

The RMSE quantifies the average magnitude of the prediction error, providing a measure of the overall accuracy of the model. The ACC evaluates the linear association between the predicted and observed anomalies, serving as an indicator of the model's skill in capturing the spatiotemporal variability of the ocean field. The Bias represents the systematic offset between predictions and observations, while ACT is defined as the standard deviation of the predicted anomalies and reflects the model's ability to reproduce the observed variability amplitude given a measure of the smoothness of the forecast.

\subsection{Anomaly Correlation Coefficient (ACC)}
The ACC measures the correlation between the predicted and observed anomalies, evaluating the model's ability to predict deviations from the climatology. It is calculated as:
\begin{equation}\label{eq:acc}
    \text{ACC} = \frac{ 
    \overline{(a_f - \overline{a_f})(a_o - \overline{a_o})}
    }
    {
    \sqrt{\overline{(a_f - \overline{a_f})^2}}\sqrt{\overline{(a_o - \overline{a_o})^2}} 
    },
\end{equation}
where $a_f = \hat{x_i^t} - c_i^t$ denotes the predicted anomaly (prediction deviation from climatology $c_i^t$) and $a_o = x_i^t - c_i^t$ the observed anomaly (ground truth deviation) for a given time and location. $\overline{(\cdot)} = \latwavr[\cdot]$ represents the latitudinal weighted spatial average, where $w_\phi$ is the latitude-weighting factor based on the cosine of the latitude expressed in radians \citep{Geer01122016,rasp2020weatherbench, lam2023graphcast}. This weighting scheme ensures that regions near the equator, where the longitudinal distance between grid points is larger, are not underrepresented in the score calculation.

The ACC ranges between -1 and 1, where values close to 1 indicate a high positive correlation between the predicted and observed anomalies. This metric is widely used in forecast verification studies (ECMWF).

\subsection{Root Mean Square Error (RMSE)}
The RMSE quantifies the average magnitude of the difference between the predicted and observed values, providing a measure of the model's overall accuracy. It is defined as:
\begin{equation}\label{eq:rmse}
    \text{RMSE} = \sqrt{\overline{\left(\hat{x}_i^t - x_i^t\right)^2}},
\end{equation}
where $\hat{x}_i^t$ is the predicted value and $x_i^t$ is the ground truth value at a given time and grid cell.

A lower RMSE indicates better predictive accuracy, making it one of the most common metrics for forecast verification.

\subsection{Activity (ACT)}
The Activity assesses the model's ability to reproduce the observed variability by measuring the standard deviation of the predicted and observed anomaly fields:
\begin{align}\label{eq:activity}
    & \text{ACT}_f = \sqrt{\overline{(a_f - \overline{a_f})^2}}, &  \text{ACT}_o = \sqrt{\overline{(a_o - \overline{a_o})^2}}.\\
\end{align}

A similar ACT between predictions and observations suggests that the model accurately captures the spatial variability of the ocean field. This score was originally proposed by~\cite{thorpe2013evaluation} and later adopted by ~\cite{RiseofDataDriven} to assess the smoothness of the forecast.

The relative activity (RA) of a model is defined as the ratio between the forecast ACT and the observed ACT~\citep{bechtold2008advances}. Analyzing RA as a function of forecast lead time reveals whether the model underestimates variability (producing smoother forecasts) or overestimates it (producing noisier forecasts):
\begin{equation}\label{eq:relative_activity}
    \text{RA} = \frac{\text{ACT}_f}{\text{ACT}_o}.
\end{equation}

\subsection{Bias}
The Bias quantifies the systematic difference between the predicted and observed values, indicating whether the model tends to overestimate or underestimate the observations. It is calculated as:
\begin{equation}\label{eq:bias}
    \text{Bias} = \overline{\left(\hat{x}_i^t - x_i^t\right)}.
\end{equation}

A Bias close to zero implies that the model does not present systematic overestimation or underestimation of the observations. This metric is commonly used in the verification of ocean and atmospheric models.

These metrics provide a comprehensive assessment of the models' performance, allowing us to quantify their predictive skill across different aspects of the forecast quality.

\bibliographystyle{apalike} 
\bibliography{references}  

\begin{thebibliography}{}

\bibitem[Adcroft et~al., 2010]{GulfMexico}
Adcroft, A., Hallberg, R., Dunne, J.~P., Samuels, B.~L., Galt, J.~A., Barker, C.~H., and Payton, D. (2010).
\newblock Simulations of underwater plumes of dissolved oil in the {Gulf of Mexico}.
\newblock {\em Geophysical Research Letters}, 37(18).

\bibitem[Arístegui et~al., 2009]{ARISTEGUI200933}
Arístegui, J., Barton, E.~D., Álvarez Salgado, X.~A., Santos, A. M.~P., Figueiras, F.~G., Kifani, S., Hernández-León, S., Mason, E., Machú, E., and Demarcq, H. (2009).
\newblock Sub-regional ecosystem variability in the {Canary Current upwelling}.
\newblock {\em Progress in Oceanography}, 83(1):33--48.
\newblock Eastern Boundary Upwelling Ecosystems: Integrative and Comparative Approaches.

\bibitem[Arístegui et~al., 1994]{ARISTEGUI19941509}
Arístegui, J., Sangrá, P., Hernández-León, S., Cantón, M., Hernández-Guerra, A., and Kerling, J. (1994).
\newblock Island-induced eddies in the {Canary Islands}.
\newblock {\em Deep Sea Research Part I: Oceanographic Research Papers}, 41(10):1509--1525.

\bibitem[Barton et~al., 1998]{BARTON1998455}
Barton, E., Arístegui, J., Tett, P., Cantón, M., García-Braun, J., Hernández-León, S., Nykjaer, L., Almeida, C., Almunia, J., Ballesteros, S., Basterretxea, G., Escánez, J., García-Weill, L., Hernández-Guerra, A., López-Laatzen, F., Molina, R., Montero, M., Navarro-Pérez, E., Rodríguez, J., {van Lenning}, K., Vélez, H., and Wild, K. (1998).
\newblock The transition zone of the {Canary Current} upwelling region.
\newblock {\em Progress in Oceanography}, 41(4):455--504.

\bibitem[{Battaglia} et~al., 2018]{2018arXiv180601261B}
{Battaglia}, P.~W., {Hamrick}, J.~B., {Bapst}, V., {Sánchez-González}, A., {Zambaldi}, V., {Malinowski}, M., {Tacchetti}, A., {Raposo}, D., {Santoro}, A., {Faulkner}, R., {Gulcehre}, C., {Song}, F., {Ballard}, A., {Gilmer}, J., {Dahl}, G., {Vaswani}, A., {Allen}, K., {Nash}, C., {Langston}, V., {Dyer}, C., {Heess}, N., {Wierstra}, D., {Kohli}, P., {Botvinick}, M., {Vinyals}, O., {Li}, Y., and {Pascanu}, R. (2018).
\newblock {Relational inductive biases, deep learning, and graph networks}.
\newblock {\em arXiv e-prints}, page arXiv:1806.01261.

\bibitem[Battaglia et~al., 2016]{battaglia2016interactionnetworkslearningobjects}
Battaglia, P.~W., Pascanu, R., Lai, M., Rezende, D., and Kavukcuoglu, K. (2016).
\newblock {Interaction Networks} for learning about objects, relations and physics.
\newblock Technical report, Cornell University.

\bibitem[Bechtold et~al., 2008]{bechtold2008advances}
Bechtold, P., K{\"o}hler, M., Jung, T., Leutbecher, M., Rodwell, M., and Vitart, F. (2008).
\newblock Advances in simulating atmospheric variability with {IFS} cycle 32r3.
\newblock {\em ECMWF Newsletter No. 114 - Winter 2007/08}, 114:29--38.

\bibitem[Bell et~al., 2009]{GODAE}
Bell, M.~J., Lefèbvre, M., Traon, P.-Y.~L., Smith, N., and Wilmer-Becker, K. (2009).
\newblock {GODAE}: the global ocean data assimilation experiment.
\newblock {\em Oceanography}, 22(3):14--21.

\bibitem[Bengio et~al., 1994]{279181}
Bengio, Y., Simard, P., and Frasconi, P. (1994).
\newblock Learning long-term dependencies with gradient descent is difficult.
\newblock {\em IEEE Transactions on Neural Networks}, 5(2):157--166.

\bibitem[Bi et~al., 2023]{Pangu}
Bi, K., Xie, L., Zhang, H., Chen, X., Gu, X., and Tian, Q. (2023).
\newblock Accurate medium-range global weather forecasting with 3d neural networks.
\newblock {\em Nature}, 619(7970):533--538.

\bibitem[Bindoff et~al., 2019]{Bindoff2019}
Bindoff, N.~L., Cheung, W. W.~L., Kairo, J.~G., Arístegui, J., Guinder, V.~A., Hallberg, R., Hilmi, N., Jiao, N., Karim, M.~S., Levin, L., O'Donoghue, S., Cuicapusa, S. R.~P., Rinkevich, B., Suga, T., Tagliabue, A., and Williamson, P. (2019).
\newblock Changing ocean, marine ecosystems, and dependent communities.
\newblock In Pörtner, H.-O., Roberts, D.~C., Masson-Delmotte, V., Zhai, P., Tignor, M., Poloczanska, E., Mintenbeck, K., Alegría, A., Nicolai, M., Okem, A., Petzold, J., Rama, B., and Weyer, N.~M., editors, {\em IPCC Special Report on the Ocean and Cryosphere in a Changing Climate}, pages 447--587. Cambridge University Press, Cambridge, UK and New York, NY, USA.

\bibitem[Bodnar et~al., 2024]{Aurora}
Bodnar, C., Bruinsma, W.~P., Lucic, A., Stanley, M., Brandstetter, J., Garvan, P., Riechert, M., Weyn, J., Dong, H., Vaughan, A., Gupta, J.~K., Tambiratnam, K., Archibald, A., Heider, E., Welling, M., Turner, R.~E., and Perdikaris, P. (2024).
\newblock A foundation model of the atmosphere.
\newblock Technical report, Cornell University.

\bibitem[Bouallègue et~al., 2024]{RiseofDataDriven}
Bouallègue, Z.~B., Clare, M. C.~A., Magnusson, L., Gascón, E., Maier-Gerber, M., Janoušek, M., Rodwell, M., Pinault, F., Dramsch, J.~S., Lang, S. T.~K., Raoult, B., Rabier, F., Chevallier, M., Sandu, I., Dueben, P., Chantry, M., and Pappenberger, F. (2024).
\newblock The rise of data-driven weather forecasting: A first statistical assessment of machine learning–based weather forecasts in an operational-like context.
\newblock {\em Bulletin of the American Meteorological Society}, 105(6):E864 -- E883.

\bibitem[Brasseur and Verron, 2006]{brasseur2006seek}
Brasseur, P. and Verron, J. (2006).
\newblock The {SEEK} filter method for data assimilation in oceanography: a synthesis.
\newblock {\em Ocean Dynamics}, 56:650--661.

\bibitem[Cabanes et~al., 2013]{os-9-1-2013}
Cabanes, C., Grouazel, A., von Schuckmann, K., Hamon, M., Turpin, V., Coatanoan, C., Paris, F., Guinehut, S., Boone, C., Ferry, N., de~Boyer~Mont\'egut, C., Carval, T., Reverdin, G., Pouliquen, S., and Le~Traon, P.-Y. (2013).
\newblock The {CORA} dataset: validation and diagnostics of in-situ ocean temperature and salinity measurements.
\newblock {\em Ocean Science}, 9(1):1--18.

\bibitem[Chambers et~al., 2017]{chambers2017evaluation}
Chambers, D.~P., Cazenave, A., Champollion, N., Dieng, H., Llovel, W., Forsberg, R., von Schuckmann, K., and Wada, Y. (2017).
\newblock Evaluation of the global mean sea level budget between 1993 and 2014.
\newblock {\em Integrative study of the mean sea level and its components}, pages 315--333.

\bibitem[Chattopadhyay et~al., 2024]{chattopadhyay_oceannet_2024}
Chattopadhyay, A., Gray, M., Wu, T., Lowe, A.~B., and He, R. (2024).
\newblock {OceanNet}: a principled neural operator-based digital twin for regional oceans.
\newblock {\em Scientific Reports}, 14(1):21181.

\bibitem[Chattopadhyay et~al., 2021]{DataDrivenForecastTurbulence}
Chattopadhyay, A., Mustafa, M., Hassanzadeh, P., and Kashinath, K. (2021).
\newblock Deep spatial transformers for autoregressive data-driven forecasting of geophysical turbulence.
\newblock In {\em Proceedings of the 10th International Conference on Climate Informatics}, CI2020, page 106–112, New York, NY, USA. Association for Computing Machinery.

\bibitem[{Chattopadhyay} et~al., 2023]{2023arXiv230407029C}
{Chattopadhyay}, A., {Sun}, Y.~Q., and {Hassanzadeh}, P. (2023).
\newblock {Challenges of learning multi-scale dynamics with {AI} weather models: Implications for stability and one solution}.
\newblock {\em arXiv e-prints}, page arXiv:2304.07029.

\bibitem[{CMEMS}, 2024]{CopernicusMarineService2023}
{CMEMS} (2024).
\newblock {European North West Shelf/Iberia Biscay Irish} seas - high resolution l4 sea surface temperature reprocessed.

\bibitem[Couairon et~al., 2024]{archesweather}
Couairon, G., Lessig, C., Charantonis, A., and Monteleoni, C. (2024).
\newblock {ArchesWeather}: An efficient {AI} weather forecasting model at 1.5$^o$ resolution.
\newblock Technical Report 2405.14527, Cornell University.

\bibitem[Cropper et~al., 2014]{CROPPER201494}
Cropper, T.~E., Hanna, E., and Bigg, G.~R. (2014).
\newblock Spatial and temporal seasonal trends in coastal upwelling off {Northwest Africa}, 1981–2012.
\newblock {\em Deep Sea Research Part I: Oceanographic Research Papers}, 86:94--111.

\bibitem[Cushman-Roisin et~al., 1990]{WestwardMotionofMesoscaleEddies}
Cushman-Roisin, B., Tang, B., and Chassignet, E.~P. (1990).
\newblock Westward motion of mesoscale eddies.
\newblock {\em Journal of Physical Oceanography}, 20(5):758 -- 768.

\bibitem[Dee et~al., 2011]{ERA-Interim}
Dee, D.~P., Uppala, S.~M., Simmons, A.~J., Berrisford, P., Poli, P., Kobayashi, S., Andrae, U., Balmaseda, M.~A., Balsamo, G., Bauer, P., Bechtold, P., Beljaars, A. C.~M., van~de Berg, L., Bidlot, J., Bormann, N., Delsol, C., Dragani, R., Fuentes, M., Geer, A.~J., Haimberger, L., Healy, S.~B., Hersbach, H., Hólm, E.~V., Isaksen, L., Kållberg, P., Köhler, M., Matricardi, M., McNally, A.~P., Monge-Sanz, B.~M., Morcrette, J.-J., Park, B.-K., Peubey, C., de~Rosnay, P., Tavolato, C., Thépaut, J.-N., and Vitart, F. (2011).
\newblock The {ERA-Interim} reanalysis: configuration and performance of the data assimilation system.
\newblock {\em Quarterly Journal of the Royal Meteorological Society}, 137(656):553--597.

\bibitem[Docquier et~al., 2019]{docquier2019impact}
Docquier, D., Grist, J.~P., Roberts, M.~J., Roberts, C.~D., Semmler, T., Ponsoni, L., Massonnet, F., Sidorenko, D., Sein, D.~V., Iovino, D., et~al. (2019).
\newblock Impact of model resolution on {Arctic} sea ice and {North Atlantic} ocean heat transport.
\newblock {\em Climate Dynamics}, 53:4989--5017.

\bibitem[Dueben and Bauer, 2018]{gmd-11-3999-2018}
Dueben, P.~D. and Bauer, P. (2018).
\newblock Challenges and design choices for global weather and climate models based on machine learning.
\newblock {\em Geoscientific Model Development}, 11(10):3999--4009.

\bibitem[Estrada-Allis et~al., 2023]{10.3389/fmars.2023.1113879}
Estrada-Allis, S.~N., Rodríguez-Santana, {\'A}., Naveira-Garabato, A.~C., García-Weil, L., Arcos-Pulido, M., and Emelianov, M. (2023).
\newblock Enhancement of turbulence and nutrient fluxes within an {Eastern Boundary Upwelling Filament}: a diapycnal entrainment approach.
\newblock {\em Frontiers in Marine Science}, Volume 10 - 2023.

\bibitem[Ezraty et~al., 2007]{Ezraty}
Ezraty, R., Girard-Ardhuin, F., Pioll{\'e}, J.-F., Kaleschke, L., and Heygster, G. (2007).
\newblock {Arctic and Antarctic} sea ice concentration and arctic sea ice drift estimated from special sensor microwave data.
\newblock {\em D{\'e}partement d’Oc{\'e}anographie Physique et Spatiale, IFREMER, Brest, France and University of Bremen Germany}, 2.

\bibitem[Falkowski et~al., 1991]{falkowski1991role}
Falkowski, P.~G., Ziemann, D., Kolber, Z., and Bienfang, P.~K. (1991).
\newblock Role of eddy pumping in enhancing primary production in the ocean.
\newblock {\em Nature}, 352(6330):55--58.

\bibitem[Galloudec et~al., 2024]{CMEMS-GLO-PUM-001-024}
Galloudec, O.~L., Chune, S.~L., Nouel, L., Fernandez, E., Derval, C., Tressol, M., Dussurget, R., Biardeau, A., and Tonani, M. (2024).
\newblock {\em Product User Manual for Global Ocean Physics Analysis and Forecasting Product}.
\newblock Copernicus Marine Environment Monitoring Service (CMEMS).
\newblock CMEMS-GLO-PUM-001-024, Issue 2.1.

\bibitem[Geer, 2016]{Geer01122016}
Geer, A.~J. (2016).
\newblock Significance of changes in medium-range forecast scores.
\newblock {\em Tellus A: Dynamic Meteorology and Oceanography}, 68(1):30229.

\bibitem[Gers and Schmidhuber, 2000]{861302}
Gers, F. and Schmidhuber, J. (2000).
\newblock Recurrent nets that time and count.
\newblock In {\em Proceedings of the IEEE-INNS-ENNS International Joint Conference on Neural Networks. IJCNN 2000. Neural Computing: New Challenges and Perspectives for the New Millennium}, volume~3, pages 189--194 vol.3.

\bibitem[Gers et~al., 1999]{818041}
Gers, F., Schmidhuber, J., and Cummins, F. (1999).
\newblock Learning to forget: continual prediction with {LSTM}.
\newblock In {\em 1999 Ninth International Conference on Artificial Neural Networks ICANN 99. (Conf. Publ. No. 470)}, volume~2, pages 850--855 vol.2.

\bibitem[Greff et~al., 2017]{7508408}
Greff, K., Srivastava, R.~K., Koutník, J., Steunebrink, B.~R., and Schmidhuber, J. (2017).
\newblock {LSTM}: A search space odyssey.
\newblock {\em IEEE Transactions on Neural Networks and Learning Systems}, 28(10):2222--2232.

\bibitem[Hagen et~al., 1996]{hagen1996near}
Hagen, E., Zulicke, C., and Feistel, R. (1996).
\newblock Near-surface structures in the {Cape Ghir} filament off {Morocco}.
\newblock {\em Oceanologica Acta}, 19(6):577--598.

\bibitem[Hausmann and Czaja, 2012]{HAUSMANN201260}
Hausmann, U. and Czaja, A. (2012).
\newblock The observed signature of mesoscale eddies in sea surface temperature and the associated heat transport.
\newblock {\em Deep Sea Research Part I: Oceanographic Research Papers}, 70:60--72.

\bibitem[Hochreiter and Schmidhuber, 1997]{lstm}
Hochreiter, S. and Schmidhuber, J. (1997).
\newblock {Long Short-Term Memory}.
\newblock {\em Neural Computation}, 9(8):1735--1780.

\bibitem[{Holmberg} et~al., 2024]{Seacast}
{Holmberg}, D., {Clementi}, E., and {Roos}, T. (2024).
\newblock {Regional Ocean Forecasting with Hierarchical Graph Neural Networks}.
\newblock {\em arXiv e-prints}, page arXiv:2410.11807.

\bibitem[Hoyer and Hamman, 2017]{hoyer2017xarray}
Hoyer, S. and Hamman, J. (2017).
\newblock xarray: {N-D} labeled arrays and datasets in {Python}.
\newblock {\em Journal of Open Research Software}, 5(1).

\bibitem[Jean-Michel et~al., 2021]{GLORYS12}
Jean-Michel, L., Eric, G., Romain, B.-B., Gilles, G., Angélique, M., Marie, D., Clément, B., Mathieu, H., Olivier, L.~G., Charly, R., Tony, C., Charles-Emmanuel, T., Florent, G., Giovanni, R., Mounir, B., Yann, D., and Pierre-Yves, L.~T. (2021).
\newblock The {Copernicus} global $1/12^\circ$ oceanic and sea ice {GLORYS12} reanalysis.
\newblock {\em Frontiers in Earth Science}, 9.

\bibitem[{Keisler}, 2022]{2022arXiv220207575K}
{Keisler}, R. (2022).
\newblock {Forecasting Global Weather with Graph Neural Networks}.
\newblock {\em arXiv e-prints}, page arXiv:2202.07575.

\bibitem[Kochkov et~al., 2023]{NeuralGCM}
Kochkov, D., Yuval, J., Langmore, I., Norgaard, P.~C., Smith, J.~A., Mooers, G., Kl{\"o}wer, M., Lottes, J., Rasp, S., D{\"u}ben, P.~D., Hatfield, S., Battaglia, P.~W., Sánchez-González, A., Willson, M., Brenner, M.~P., and Hoyer, S. (2023).
\newblock Neural general circulation models for weather and climate.
\newblock {\em Nature}, 632:1060 -- 1066.

\bibitem[Lam et~al., 2023]{lam2023graphcast}
Lam, R., Sánchez-González, A., Willson, M., Wirnsberger, P., Fortunato, M., Alet, F., Ravuri, S., Ewalds, T., Eaton-Rosen, Z., Hu, W., Merose, A., Hoyer, S., Holland, G., Vinyals, O., Stott, J., Pritzel, A., Mohamed, S., and Battaglia, P. (2023).
\newblock Learning skillful medium-range global weather forecasting.
\newblock {\em Science}, 382(6677):1416--1421.

\bibitem[{Lang} et~al., 2024]{2024arXiv240601465L}
{Lang}, S., {Alexe}, M., {Chantry}, M., {Dramsch}, J., {Pinault}, F., {Raoult}, B., {Clare}, M. C.~A., {Lessig}, C., {Maier-Gerber}, M., {Magnusson}, L., {Ben Bouall{\`e}gue}, Z., {Prieto Nemesio}, A., {Dueben}, P.~D., {Brown}, A., {Pappenberger}, F., and {Rabier}, F. (2024).
\newblock {{AIFS -- ECMWF}'s data-driven forecasting system}.
\newblock {\em arXiv e-prints}, page arXiv:2406.01465.

\bibitem[Lellouche et~al., 2018]{os-14-1093-2018}
Lellouche, J.-M., Greiner, E., Le~Galloudec, O., Garric, G., Regnier, C., Drevillon, M., Benkiran, M., Testut, C.-E., Bourdalle-Badie, R., Gasparin, F., Hernandez, O., Levier, B., Drillet, Y., Remy, E., and Le~Traon, P.-Y. (2018).
\newblock Recent updates to the {Copernicus Marine Service} global ocean monitoring and forecasting real-time $1/12^\circ$ high-resolution system.
\newblock {\em Ocean Science}, 14(5):1093--1126.

\bibitem[Lellouche et~al., 2013]{os-9-57-2013}
Lellouche, J.-M., Le~Galloudec, O., Dr\'evillon, M., R\'egnier, C., Greiner, E., Garric, G., Ferry, N., Desportes, C., Testut, C.-E., Bricaud, C., Bourdall\'e-Badie, R., Tranchant, B., Benkiran, M., Drillet, Y., Daudin, A., and De~Nicola, C. (2013).
\newblock Evaluation of global monitoring and forecasting systems at {Mercator Océan}.
\newblock {\em Ocean Science}, 9(1):57--81.

\bibitem[{Li} et~al., 2020]{2020arXiv200615704L}
{Li}, S., {Zhao}, Y., {Varma}, R., {Salpekar}, O., {Noordhuis}, P., {Li}, T., {Paszke}, A., {Smith}, J., {Vaughan}, B., {Damania}, P., and {Chintala}, S. (2020).
\newblock {{PyTorch} Distributed: Experiences on Accelerating Data Parallel Training}.
\newblock {\em arXiv e-prints}, page arXiv:2006.15704.

\bibitem[Liu et~al., 2013]{liu2013monitoring}
Liu, Y., MacFadyen, A., Ji, Z.-G., and Weisberg, R.~H. (2013).
\newblock {\em Monitoring and modeling the deepwater horizon oil spill: a record breaking enterprise}.
\newblock John Wiley \& Sons.

\bibitem[Madec et~al., 2024]{NEMO}
Madec, G., Bell, M., Benshila, R., Blaker, A., Boudrallé-Badie, R., Bricaud, C., Bruciaferri, D., Carneiro, D., Castrillo, M., Calvert, D., Chanut, J., Clementi, E., Coward, A., de~Lavergne, C., Dobricic, S., Epicoco, I., Éthé, C., Fiedler, E., Ford, D., Furner, R., Ganderton, J., Graham, T., Harle, J., Hutchinson, K., Iovino, D., King, R., Lea, D., Levy, C., Lovato, T., Maisonnave, E., Mak, J., Sánchez, J. M.~C., Martin, M., Martin, N., Martins, D., Masson, S., Mathiot, P., Mele, F., Mocavero, S., Moulin, A., Müller, S., Nurser, G., Oddo, P., Paronuzzi, S., Paul, J., Peltier, M., Person, R., Rousset, C., Rynders, S., Samson, G., Schroeder, D., Storkey, D., Storto, A., Téchené, S., Vancoppenolle, M., and Wilson, C. (2024).
\newblock Nemo ocean engine reference manual.

\bibitem[Madec et~al., 2008]{madec2008nemo}
Madec, G. et~al. (2008).
\newblock {NEMO} ocean engine. note du {P{\^o}le} de mod{\'e}lisation.
\newblock {\em Institut Pierre-Simon Laplace (IPSL), France}, 27:1288--1619.

\bibitem[Mourre et~al., 2018]{Mourre2018AssessmentOH}
Mourre, B., Aguiar, E., Juz{\`a}, M., Hern{\'a}ndez-Lasheras, J., Reyes, E., Heslop, E., Escudier, R., Cutolo, E., Ruiz, S., Mason, E., Pascual, A., and Tintor{\'e}, J. (2018).
\newblock Assessment of high-resolution regional ocean prediction systems using multi-platform observations: Illustrations in the {Western Mediterranean Sea}.
\newblock {\em New Frontiers in Operational Oceanography}.

\bibitem[Mu et~al., 2019]{8851967}
Mu, B., Peng, C., Yuan, S., and Chen, L. (2019).
\newblock {ENSO} forecasting over multiple time horizons using {ConvLSTM} network and rolling mechanism.
\newblock In {\em 2019 International Joint Conference on Neural Networks (IJCNN)}, pages 1--8.

\bibitem[Mu et~al., 2021]{gmd-14-6977-2021}
Mu, B., Qin, B., and Yuan, S. (2021).
\newblock {ENSO-ASC} 1.0.0: {ENSO} deep learning forecast model with a multivariate air--sea coupler.
\newblock {\em Geoscientific Model Development}, 14(11):6977--6999.

\bibitem[{Oskarsson} et~al., 2023]{oskarsson2023graphbased}
{Oskarsson}, J., {Landelius}, T., and {Lindsten}, F. (2023).
\newblock {Graph-based Neural Weather Prediction for Limited Area Modeling}.
\newblock {\em arXiv e-prints}, page arXiv:2309.17370.

\bibitem[Pelegrí et~al., 2005]{PELEGRI20053}
Pelegrí, J., Arístegui, J., Cana, L., González-Dávila, M., Hernández-Guerra, A., Hernández-León, S., Marrero-Díaz, A., Montero, M., Sangrà, P., and Santana-Casiano, M. (2005).
\newblock Coupling between the open ocean and the coastal upwelling region off {Northwest Africa}: water recirculation and offshore pumping of organic matter.
\newblock {\em Journal of Marine Systems}, 54(1):3--37.
\newblock A general study of the Spanish North Atlantic boundaries: an interdisciplinary approach.

\bibitem[{Pfaff} et~al., 2020]{2020arXiv201003409P}
{Pfaff}, T., {Fortunato}, M., {Sánchez-González}, A., and {Battaglia}, P.~W. (2020).
\newblock {Learning Mesh-Based Simulation with Graph Networks}.
\newblock {\em arXiv e-prints}, page arXiv:2010.03409.

\bibitem[Pioll{\'e} and Autret, 2023]{piolleQUIDSSTTAC2023}
Pioll{\'e}, J.-F. and Autret, E. (2023).
\newblock {{QUID}} for {{SST TAC Products SST}}\_{{GLO}}\_{{SST}}\_{{L3S}}\_{{NRT}}\_{{OBSERVATIONS}}\_010\_010.
\newblock Quality Information Document 2.1, E.U. Copernicus Marine Service Information (CMEMS).

\bibitem[Price et~al., 2024]{Gencast}
Price, I., Sánchez-González, A., Alet, F., Andersson, T.~R., El-Kadi, A., Masters, D., Ewalds, T., Stott, J., Mohamed, S., Battaglia, P.~W., Lam, R., and Willson, M. (2024).
\newblock Probabilistic weather forecasting with machine learning.
\newblock {\em Nature}, 637:84 -- 90.

\bibitem[Pujol et~al., 2016]{os-12-1067-2016}
Pujol, M.-I., Faug\`ere, Y., Taburet, G., Dupuy, S., Pelloquin, C., Ablain, M., and Picot, N. (2016).
\newblock {DUACS DT2014}: the new multi-mission altimeter data set reprocessed over 20 years.
\newblock {\em Ocean Science}, 12(5):1067--1090.

\bibitem[Rasp et~al., 2020]{rasp2020weatherbench}
Rasp, S., Dueben, P.~D., Scher, S., Weyn, J.~A., Mouatadid, S., and Thuerey, N. (2020).
\newblock {WeatherBench}: a benchmark data set for data-driven weather forecasting.
\newblock {\em Journal of Advances in Modeling Earth Systems}, 12(11):e2020MS002203.

\bibitem[Rasp et~al., 2024]{rasp2024weatherbench}
Rasp, S., Hoyer, S., Merose, A., Langmore, I., Battaglia, P., Russel, T., Sánchez-González, A., Yang, V., Carver, R., Agrawal, S., Chantry, M., Bouallegue, Z.~B., Dueben, P., Bromberg, C., Sisk, J., Barrington, L., Bell, A., and Sha, F. (2024).
\newblock {WeatherBench} 2: A benchmark for the next generation of data-driven global weather models.
\newblock Technical Report 2308.15560, Cornell University.

\bibitem[Rio et~al., 2011]{https://doi.org/10.1029/2010JC006505}
Rio, M.~H., Guinehut, S., and Larnicol, G. (2011).
\newblock New {CNES-CLS09} global mean dynamic topography computed from the combination of {GRACE} data, altimetry, and in situ measurements.
\newblock {\em Journal of Geophysical Research: Oceans}, 116(C7).

\bibitem[Roberts et~al., 2018]{roberts2018benefits}
Roberts, M., Vidale, P., Senior, C., Hewitt, H., Bates, C., Berthou, S., Chang, P., Christensen, H., Danilov, S., Demory, M.-E., et~al. (2018).
\newblock The benefits of global high resolution for climate simulation: process understanding and the enabling of stakeholder decisions at the regional scale.
\newblock {\em Bulletin of the American Meteorological Society}, 99(11):2341--2359.

\bibitem[Sangrà et~al., 2009]{SANGRA20092100}
Sangrà, P., Pascual, A., Ángel Rodríguez-Santana, Machín, F., Mason, E., McWilliams, J.~C., Pelegrí, J.~L., Dong, C., Rubio, A., Arístegui, J., Ángeles Marrero-Díaz, Hernández-Guerra, A., Martínez-Marrero, A., and Auladell, M. (2009).
\newblock The {Canary Eddy Corridor}: A major pathway for long-lived eddies in the subtropical {North Atlantic}.
\newblock {\em Deep Sea Research Part I: Oceanographic Research Papers}, 56(12):2100--2114.

\bibitem[Sangrà et~al., 2015]{cape_ghir_sangra}
Sangrà, P., Troupin, C., Barreiro-González, B., Desmond~Barton, E., Orbi, A., and Arístegui, J. (2015).
\newblock The cape {G}hir filament system in {A}ugust 2009 ({NW A}frica).
\newblock {\em Journal of Geophysical Research: Oceans}, 120(6):4516--4533.

\bibitem[Scher, 2018]{Scher2018}
Scher, S. (2018).
\newblock Toward data-driven weather and climate forecasting: Approximating a simple general circulation model with deep learning.
\newblock {\em Geophysical Research Letters}, 45(22):12,616--12,622.

\bibitem[{Shi} et~al., 2015]{2015arXiv150604214S}
{Shi}, X., {Chen}, Z., {Wang}, H., {Yeung}, D.-Y., {Wong}, W.-k., and {Woo}, W.-c. (2015).
\newblock {Convolutional {LSTM} Network: A Machine Learning Approach for Precipitation Nowcasting}.
\newblock {\em arXiv e-prints}, page arXiv:1506.04214.

\bibitem[Sommer et~al., 2018]{Sommer2018OceanCM}
Sommer, J.~L., Chassignet, E.~P., and Wallcraft, A.~J. (2018).
\newblock Ocean circulation modeling for operational oceanography: Current status and future challenges.
\newblock {\em New Frontiers in Operational Oceanography}.

\bibitem[Szekely et~al., 2019]{szekely2019cora}
Szekely, T., Gourrion, J., Pouliquen, S., Reverdin, G., and Merceur, F. (2019).
\newblock {CORA}, coriolis ocean dataset for reanalysis.
\newblock {\em Sea Scientific Open Data Publication}.

\bibitem[Thompson et~al., 2021]{gmd-14-1081-2021}
Thompson, B., Sánchez, C., Heng, B. C.~P., Kumar, R., Liu, J., Huang, X.-Y., and Tkalich, P. (2021).
\newblock Development of a {MetUM} (v 11.1) and {NEMO} (v 3.6) coupled operational forecast model for the maritime continent -- {Part} 1: Evaluation of ocean forecasts.
\newblock {\em Geoscientific Model Development}, 14(2):1081--1100.

\bibitem[Thorpe et~al., 2013]{thorpe2013evaluation}
Thorpe, A., Bauer, P., Magnusson, L., and Richardson, D. (2013).
\newblock An evaluation of recent performance of {ECMWF}’s forecasts.
\newblock {\em ECMWF Newsletter No. 137 - autumn 2013}, pages 15--18.

\bibitem[Treguier et~al., 2017]{Treguier2017ModelingAF}
Treguier, A.-M., Chassignet, E.~p., Le~Boyer, A., and Pinardi, N. (2017).
\newblock Modeling and forecasting the "weather of the ocean" at the mesoscale.
\newblock {\em Journal Of Marine Research}, 75(3):301--329.

\bibitem[Troupin et~al., 2012]{TROUPIN20121}
Troupin, C., Mason, E., Beckers, J.-M., and Sangrà, P. (2012).
\newblock Generation of the {Cape Ghir} upwelling filament: A numerical study.
\newblock {\em Ocean Modelling}, 41:1--15.

\bibitem[Wang et~al., 2024]{wangXiHeDataDrivenModel2024}
Wang, X., Wang, R., Hu, N., Wang, P., Huo, P., Wang, G., Wang, H., Wang, S., Zhu, J., Xu, J., Yin, J., Bao, S., Luo, C., Zu, Z., Han, Y., Zhang, W., Ren, K., Deng, K., and Song, J. (2024).
\newblock {{XiHe}}: {{A Data-Driven Model}} for {{Global Ocean Eddy-Resolving Forecasting}}.
\newblock Technical Report arXiv:2402.02995, Cornell University.

\bibitem[{Watters} et~al., 2017]{2017arXiv170601433W}
{Watters}, N., {Tacchetti}, A., {Weber}, T., {Pascanu}, R., {Battaglia}, P., and {Zoran}, D. (2017).
\newblock {{Visual Interaction Networks}}.
\newblock {\em arXiv e-prints}, page arXiv:1706.01433.

\bibitem[Wooster et~al., 1976]{wooster1976seasonal}
Wooster, W.~S., Bakun, A., and McLain, D.~R. (1976).
\newblock The seasonal upwelling cycle along the eastern boundary of the {N}orth {A}tlantic.
\newblock {\em Journal of Marine Research}, 34(2):131--141.

\bibitem[Yang et~al., 2024]{10.3389/fmars.2024.1470320}
Yang, J., Zhang, T., Zhang, J., Lin, X., Wang, H., and Feng, T. (2024).
\newblock A {ConvLSTM} nearshore water level prediction model with integrated attention mechanism.
\newblock {\em Frontiers in Marine Science}, 11.

\bibitem[Yang et~al., 2022]{YANG2022109003}
Yang, X., Zhang, F., Sun, P., Li, X., Du, Z., and Liu, R. (2022).
\newblock A spatio-temporal graph-guided convolutional {LSTM} for tropical cyclones precipitation nowcasting.
\newblock {\em Applied Soft Computing}, 124:109003.

\bibitem[Zeng et~al., 2015]{PredictabilityEddyShedding}
Zeng, X., Li, Y., and He, R. (2015).
\newblock Predictability of the loop current variation and eddy shedding process in the {Gulf of Mexico} using an artificial neural network approach.
\newblock {\em Journal of Atmospheric and Oceanic Technology}, 32(5):1098 -- 1111.

\end{thebibliography}

\end{document}